\documentclass[pre,showpacs,preprintnumbers,amsmath,amssymb,superscriptaddress]{revtex4}

\usepackage{graphicx}
\usepackage{bm}
\usepackage{longtable}

\usepackage[utf8]{inputenc}
\usepackage[T1]{fontenc}
\usepackage{multirow}

\begin{document}

\title{Rare events in generalized L\'evy Walks  and the Big Jump principle}

\author{}
\affiliation{}

\author{Alessandro Vezzani}
\affiliation{IMEM, CNR Parco Area delle Scienze 37/A	43124 Parma}
\affiliation{Dipartimento di Matematica, Fisica e Informatica Universit\`a degli Studi di
Parma, viale G.P.Usberti 7/A, 43124 Parma, Italy}
\author{Eli Barkai}
\affiliation{Department of Physics, Institute of Nanotechnology and Advanced Materials, Bar-Ilan University, Ramat-Gan, 52900, Israel}
\author{Raffaella Burioni}
\affiliation{
Dipartimento di Matematica, Fisica e Informatica Universit\`a degli Studi di
Parma, viale G.P.Usberti 7/A, 43124 Parma, Italy}
\affiliation{INFN, Gruppo Collegato di
Parma, viale G.P. Usberti 7/A, 43124 Parma, Italy}


\begin{abstract}
The prediction and control of rare events is an important task in disciplines that range from physics and biology, to economics and social science.  The Big Jump principle deals with a peculiar aspect of the mechanism that drives rare events. According to the principle, in heavy-tailed processes  a rare huge fluctuation is caused by a single event and not by the usual coherent accumulation of small deviations. We consider generalized L\'evy walks, a class of stochastic processes with power law distributed step durations, which model complex microscopic dynamics in the single stretch. We derive the bulk of the probability distribution and using the big jump principle, the exact form of the tails that describes rare events. We show that the tails of the distribution present non-universal and non-analytic behaviors, which  depend crucially on the dynamics of the single step. The big jump estimate also provides a physical explanation of the processes driving the rare events, opening new possibilities for their correct prediction.
\end{abstract}

\flushbottom
\maketitle
%
%
\thispagestyle{empty}


\section*{Introduction}

Rare events are an important and exciting theoretical research field in mathematics and in natural sciences, with
a long history in topics ranging from physics, geophysics and biology, to ecology and social systems \cite{Gumbel,Hollander,Holger,Vulp}. A deeper understanding
of the mechanism that leads to rare events is a major problem in risk predictions and management, across different disciplines \cite{Embrechts,Lucilla1}. 

In this field, an interesting role is played by the so called  {\it Big Jump principle}. The principle explains extreme events in a wide class of natural and man-made systems
with heavy tailed distributions, not in terms of an accumulation of many small subevents but solely as an effect of the biggest event, the big jump.  The 
accumulated rain fall in one month in a region \cite{rain}, the energy released in an earthquake,  and also the position of particles whose motion is determined by a sum of
very heterogeneous steps are examples of processes where the principle is very likely to be valid.  If only one event is controlling the statistics of extremes, we can understand the inherent difficulties
in the prediction. At the same time, if we know that the process we are studying follows the principle, we can learn how to better quantify the extremes.

The big jump principle has originally been shown to hold for sums of independent random variables following a heavy-tailed (i.e. subexponential) distribution \cite{Chistyakov,Foss,Denisov,Geluk,Clusel1}. Recently it has been applied to models of anomalous transport in  quenched disorder \cite{levyrand,Ub,VBB19}, where it has been used to predict with a surprising accuracy large fluctuations driven by a single rare event. Interestingly, a key feature of the big jump approach is that it is able to reproduce the whole general shape of the probability density for rare events, and in particular its non-analytical behaviors \cite{VBB19}, i.e. cusps and fine structures related to the specific form of the single process that contributes to the tail. Indeed, while the central part of the probability distribution typically features universal and smooth shapes driven by central limit theorems arguments \cite{Gardiner},  the big jump can give rise to non universal effects
since it involves a single process. The non universal effects can be used in one direction, from the microscopic modelling towards an accurate prediction of the risk for rare events, but also in reverse order, that is to argue details of the microscopic underlying processes from the structure of the far tail.
 
In particular, the big jump principle was recently extended \cite{VBB19,WVBB19,Gradenigo} to case studies which involve L\'evy walks. These are introduced as continuous time stochastic process for particles performing steps with duration drawn from a power law, hence heavy-tailed, distribution \cite{Klafter1,zumofen,zaburdaev}. Because of their generality, L\'evy walks are applied to describe motion of cold atoms in laser cooling \cite{Davidson}, transport in turbulent flow \cite{boffetta} and in neural transmission  \cite{Neur-levy}, animal motion \cite{Ariel,future-levy}, and natural and optimized search processes \cite{Benichou}. These systems all have in common a power law distribution for step durations and they can differ in how the walker moves along the steps,  i.e. at constant velocity or with a more complex type of motion. In this framework, the typical quantity of interest is the particle position at fixed time, independently of the number of steps (draws). This introduces a non trivial coupling mechanism between position and observation time, as the far tails of the position distribution are naturally cutoff by the finite speed of propagation. Therefore L\'evy walks depart from the simple case of summation of random variables and in particular the distribution of rare events presents cutoffs and other non analytic features.

A generalized L\'evy walks \cite{Albers,Sokolov}, originally introduced for motion in turbulent fluids, 
has recently been considered to model complex motion in each single stretch. More precisely, the duration of a step $t$ is drawn from a power law distribution $\lambda(t)\sim t^{-1-\alpha}$, while the motion within a step is described by two further exponents: $\nu$, relating the step length with the duration time $t$, and $\eta$,  which provides the temporal dynamics within a step, modelling acceleration and deceleration effects. Such a general description of the microscopic motion is suitable to deal with a wide class of L\'evy walks and hence it can be applied to many model systems in the presence of complex trajectories  \cite{Ariel,future-levy}. Previous results \cite{Albers,Sokolov} focus on the calculation of the mean square displacement of the generalized L\'evy walk as a function of time. Here, we describe the asymptotic time evolution of the entire walker probability distribution, which allows us to extract the behavior of correlations and higher moments.

First, we apply standard techniques in random walk theory to obtain the central part of the distribution and its scaling length. 
We show that the bulk of the distribution displays standard universal behaviors, i.e. a Gaussian distribution, a L\'evy stable distribution or the distribution of continuous time random walks (CTRW), depending on the divergence or finiteness of the mean duration and the mean square length of the single step. 

Then, by using the big jump principle, we characterize the tail of the probability distribution at distances much larger than the scaling length. We show that rare events are described by non trivial functions, determined  both by the duration distribution of the steps $\lambda(t)$ and by the microscopic acceleration and deceleration along the step, so that the result depends on all the exponents $\alpha$, $\nu$ and $\eta$. Remarkably, these non-universal distributions, which display non-analytic behaviors, are obtained from the general principle of single big jump, which provides a unique physical explanation of the process driving the rare events.  We also highlight that for some values of  $\alpha$, $\nu$ and $\eta$  the motion within a step can be slower than the growth of the scaling length, so in this case the principle does not apply.
As a final result, we also derive the scaling of all the moments of the distribution that, interestingly, feature strong anomalous diffusion \cite{castiglione,Cagnetta,vollmer}. All our analytical results are in very good agreement with extensive numerical simulations.
 
The paper is organized as follows: the section Results is divided into 4 parts. In the first one we discuss the single big jump principle. In the second, we discuss the generalized  L\'evy walk model \cite{Albers,Sokolov} and we describe the central part  of the probability distribution. In the third part we apply the big jump principle to the generalized L\'evy walk and we obtain the distribution of rare events and in the last part  we discuss the moments of the distribution. Comparisons with numerical simulations are shown along the sections, showing a very good agreement in the long time asymptotic limit. The section Methods is devoted to a discussion of the big jump principle in terms of a very general formulation which can be applied to a wide class of models. In the Supplementary Information (SI) we describe some 
 the details of our calculation.  We end with our conclusions and final remarks.

\section*{Results}

\subsubsection{The Big Jump principle}

The big jump principle applies to systems where a rare fluctuation of a stochastic variable is driven by a single extreme event, that we call the big jump. We introduce the principle
with the {\em rate approach} \cite{VBB19}, an heuristic formulation which allows for an easy extension beyond the standard case of sum of independent and identically distributed random variables. The estimate is based on the splitting of the problem in two parts: the first one leads to the calculation of the {jump rate}, that is the rate at which the walker makes attempts to perform the big jump. The second part takes into account the dynamical evolution during the big jump.

We consider a dynamical stochastic process with random variables $t_i$ drawn from a broad distribution $\lambda(t)$ at times $T_i$ ($T_i<T_j$ if $i<j$). The extraction time $T_i$ is also a random variable that can depend on $t_i$ ($i<j$), while in the simple case of the sum of IID we simply have $T_i=i$. We are interested in the "global" stochastic variable $R$ which in general depends on $t_i$ in a non trivial way ($R>0$ for the sake of simplicity). We call $P(R,T)$ the Probability Density Function (PDF) of measuring $R$ at time $T$ (see Methods for details). We focus on generalized L\'evy walks where the event $i$ is a jump, $t_i$ is the jump duration and $R$ the particle position. In different stochastic processes, $t_i$ and $R$ can have different interpretations (energies, masses...) \cite{Maj1b,Maj2,filias,corberi}.  

We consider a process where, at large $T$, $P(R,T)$ can be split in two terms, one related to
the central part of the distribution, describing typical values of the final position R, and the other related to  the far tail at very large $R$, driven by rare events:
\begin{equation}
P(R,T)\sim
\left\{
\begin{array}{ll}
\displaystyle
\ell^{-1}(T)f(R/\ell(T)) &\mbox{if }\ R < \ell(T)\kappa(T) \\
\displaystyle
B(R,T) &\mbox{if }\ R> \ell(T) \kappa(T) 
\end{array}
\right.
\label{sal}
\end{equation}
where $\ell(T)$ is the characteristic length of the process and $\kappa(T)$ is a slowly growing function of $T$ (e.g. a logarithmic function).
Notice that at large $T$, $P(R,T)$ converges in probability to $\ell^{-1}(T)f(R/\ell(T))$, 
a function which is significantly different from zero only for values of the final position $R < \ell(T) \kappa(T)$.
However, $B(R,T)$ describes $P(R,T)$ for $R\gg \ell(T)$, i.e. at distances much larger than the scaling length of the process. Therefore $B(R,T)$ can be relevant in the calculation of higher moments of the distribution $\langle R^q (T)\rangle=\int_0^\infty P(R,T)R^q dR$ ($q>0$), such as the mean square displacement $q=2$, since:
\begin{equation}
\langle R^q (T)\rangle \sim  \int_0^{\ell(T)\kappa(T)} \ell^{-1}(T)f(R/\ell(T)) R^q dR + \int_{\ell(T)\kappa(T)}^\infty B(R,T) R^q dR.
\label{rP}
\end{equation} 
Here the first term can be subleading with respect to the second integral for $q >q_c$, where $q_c$ is a critical order of the moments. This means that some moments of the process are influenced by the rare events \cite{castiglione,vollmer}. $B(R,T)$ is precisely the part of the distribution that we want to calculate with the big jump principle. In practice, $B(R,T)$ describes the finite time deviations of $P(R,T)$ from the bulk scaling function at large $R\gg \ell(T)$, and this is what determines the anomalous moments of the distribution. 

Since $\lambda(t)$ does not depend on the jumping time, 
the probability to perform a jump of duration $t$ at time $T_w$ is $p_{{\rm tot}}(t,T_w)=  n_R(T_w)\cdot \lambda(t) $, where  $n_R(T_w)$ is the jumping rate that is  $n_R(T)=d \langle N(T) \rangle /d T$
and $\langle N(T) \rangle $ in the average number of jumps up to time $T$.
As  we are considering $R\gg \ell(T)$, according to the principle we suppose that the only important process that contributes to $B(R,T)$  is the biggest jump and therefore we neglect all the jumps occurring before and after that. We call ${\cal P}(R|T,t,T_w)$ the probability that a process, driven by the single jump of duration $t$ starting at $T_w$, takes the walker in $R$ at time $T\geq T_w$.  The big jump principle states that, as for $R\gg \ell(T)$ the relevant part of the distribution is $B(R,T)$, this can be determined as:
\begin{equation}
B(R,T) = \int dt \int_0^T dT_w p_{{\rm tot}}(t,T_w) {\cal P}(R|T,t,T_w)
\label{BigJump}
\end{equation}
Hence, $B(R,T)$ is evaluated by summing over all the paths ($t$ and $T_w$) that in a single jump  bring the process to $R$ at time $T$. These paths, described by ${\cal P}(R|T,t,T_w)$, can be very complex, as they include all the correlations and non-linearities of the the model. However, since only one stochastic draw is involved, an analytic approach is often feasible (for further details see Methods). 
Notice that  Eq. (\ref{BigJump}) provides an estimate of $B(R,T)$ only for large $R$, so in general $B(R,T)$ can behave as an {\em infinite density} \cite{Eli1}, i.e. $B(R,T)$ diverges at $R=0$ so that $\int dR B(R,T)=\infty$. Nevertheless, $B(R,T)$ provides the correct expression for the asymptotic behavior of the moments $\langle R^q(T) \rangle$ with large $q$, since,  according to Eq. (\ref{rP}),  the factor $R^q$ cures the divergence in $R=0$. Notice also that the hypothesis that a single big jump contributes to $B(R,T)$ is crucial. If in a process it is not possible to  reach $R>>\ell(T)$
with a single stochastic event,  Eq. (\ref{BigJump}) does not apply and different approaches must be introduced  \cite{Hoell}. 

\subsubsection{Generalized L\'evy walks: microscopic dynamics and the bulk of the distribution}

The generalized L\'evy walk \cite{Albers,Sokolov}  is a model of anomalous transport with acceleration and deceleration along the microscopic trajectories, an effect that is often encountered in experiments \cite{Ariel,future-levy}.
In this model, the stochastic variable $t_i$ drawn from the broad PDF $\lambda(t_i)$ defines the duration of the $i$-th step so that the draw $i$ occurs at time $T_i=\sum_{j=1}^{i-1} t_j $ ($T_1=0$).
As a typical example of broad distribution we take a power law $\lambda(t)$ where for  $t>\tau_0$   
\begin{equation}
\lambda(t) = \frac{\tau_0^\alpha}  {t^{1 + \alpha} } 
\label{lambdaL}
\end{equation}
and $\lambda(t)=0$ for $t<\tau_0$.
We define $r(T)$ the position of the walker and $R(T)=|r(T)-r(0)|$ its distance from the origin.
The microscopic dynamic of the walker in the time interval $T_i<T<T_{i+1}$ is defined as:
\begin{equation}
r(T)=r(T_i)+ c_i t_i^{\nu-\eta} (T-T_i)^\eta, 
\label{ap2}
\end{equation}
where $\nu>0$ and $\eta>0$ are the parameters describing the microscopic motion and the random "velocity" $c_i=\pm c$ is drawn with probability $1/2$ in each step. According to Eq.(\ref{ap2}) the step $i$ starts in $r(T_i)$ and stops in $r(T_i)+c_i t_i^\nu=r(T_{i+1})$ which defines the starting point of the $i+1$ step. In this framework we call $L_i=c t_i^\nu$ the length of the step $i$.

\begin{figure}
\centering
\includegraphics[width=0.70\textwidth]{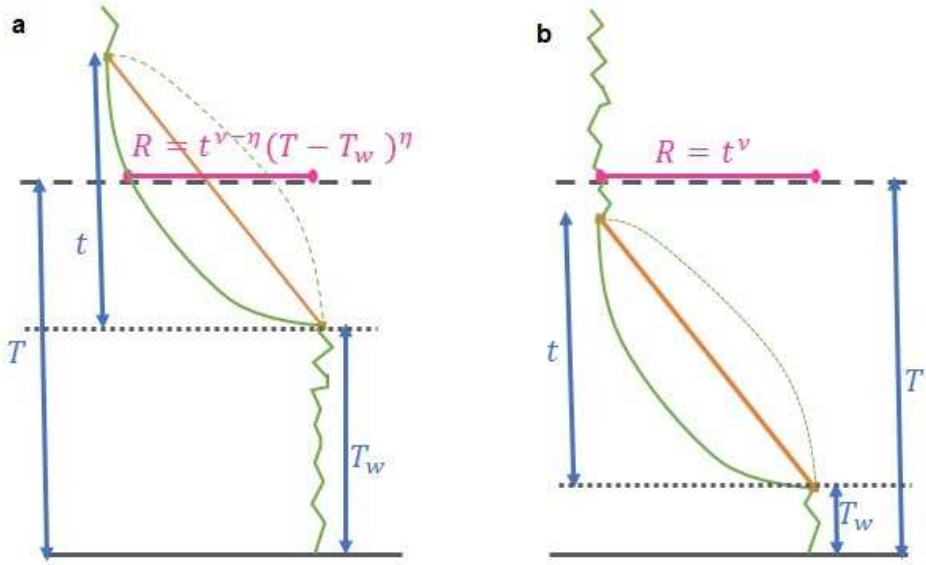}
	\caption{
		The big jump contributions.
		The jump starts at time $T_w$ and it can either lead you to the
		time horizon of the L\'evy walk $t>(T-T_w)$, as in panel ({\bf a}) or it may start and end before time $T$ if $t<(T-T_w)$ as in panel ({\bf b}). The orange line in the big jump represents the global motion after a jump $c t^\nu$ ($\nu=1$ in this case). In green we plot the motion of the walker. Within the jump we plot $c t^{\nu-\eta}(T-T_w)^\eta$ and continuous and dashed lines refer to $\eta<\nu$ and to $\eta>\nu$ respectively. 
 The final position $R$ is plotted in magenta. In panel ({\bf a}) $R$ depends on $\eta$ and $\nu$ while in panel ({\bf b}) it is driven by the exponent $\nu$ only.}
	\label{LWfig}
\end{figure}

The generalized L\'evy walks correspond to many different types of motions along the steps. If $\eta < \nu$ the walker moves faster at the beginning of the step, then it slows down. Conversely, for $\eta>\nu$ the motion starts at slow speed, then it speeds up (see Figure \ref{LWfig}). In particular, for $\eta=0$ we recover the so called step-first dynamics \cite{zaburdaev}, where the particle reaches instantaneously $r(T_i)+ c_i t_i^\nu$ at time $T_i$ then it waits a time $t_i$ before the following step. On the other hand, for $\eta=\infty$ this is the wait-first dynamics \cite{zaburdaev}, with the walker waiting a time $t_i$ in $r(T_i)$ then suddenly moving to $r(T_i)+ c_i t_i^\nu$ just before the next step.
The case $\eta=\nu=1$ corresponds to standard L\'evy walks \cite{Klafter1}, which presents ballistic motion along the steps, while the case $\eta=\nu$ has  been studied recently
in detail in \cite{Aghion}, where the distribution of the rare events has been evaluated using  a moment resummation technique.

If $\lambda(t_i)$ is given by Eq. (\ref{lambdaL}), the step length $L_i$ is distributed as  $\tilde \lambda(L_i)\sim L_i^{-1-\alpha/\nu}$.
Hereafter, we define $\langle t \rangle =\int dt t \lambda(t)$ the average duration of a step and 
$\langle L^2 \rangle=c^2 \langle t^{2\nu} \rangle= \int dt t^{2\nu} \lambda(t)$ the average square length of a jump; $\langle t \rangle$ is finite for $\alpha>1$, $\langle L^2 \rangle$ is finite for $\alpha>2\nu$. 
Since at the end of the jump the length $L_i=c t_i ^\nu$ is independent of $\eta$, one can expect 
naively, as in standard transport theories, that the statistical properties of $R$ will be $\eta$ independent. However, for heavy-tailed processes, the dynamics in the time interval between the last jump and the measurement time are important and hence the final result will be sensitive to $\eta$.

Since $t_i$ can be arbitrary large, also the generalized velocity $c t_i^{\nu-\eta}$ for $\nu>\eta$ is unbounded and the walker can reach arbitrary large distances in an arbitrary small time $\delta t=T-T_i$.
Conversely, for $\eta\geq \nu$, in a time $T$  the walker can reach a maximum distance $l_{\rm cone}(T)$, that we call the {\em light cone} of the walker. For $\nu\geq 1$ the light cone can be reached in a single step and $l_{\rm cone}(T)=c T^{\nu}$. For $\nu<1$ the light cone can only be reached in many steps, all in the same direction and $l_{\rm cone}(T)=T c \tau_0^{\nu-1}$.

\begin{figure}
\centering
\includegraphics[width=0.48\textwidth]{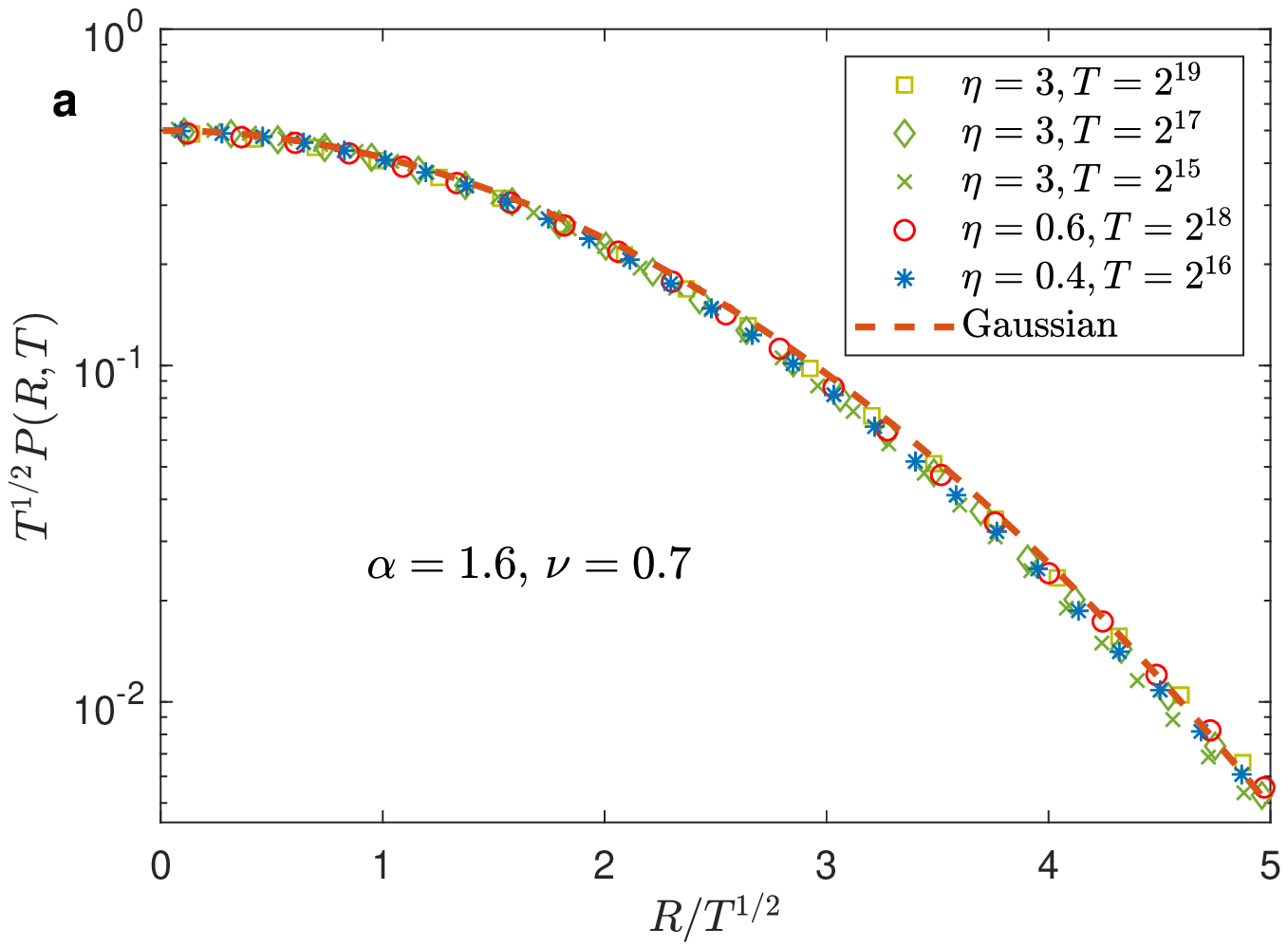}
\includegraphics[width=0.48\textwidth]{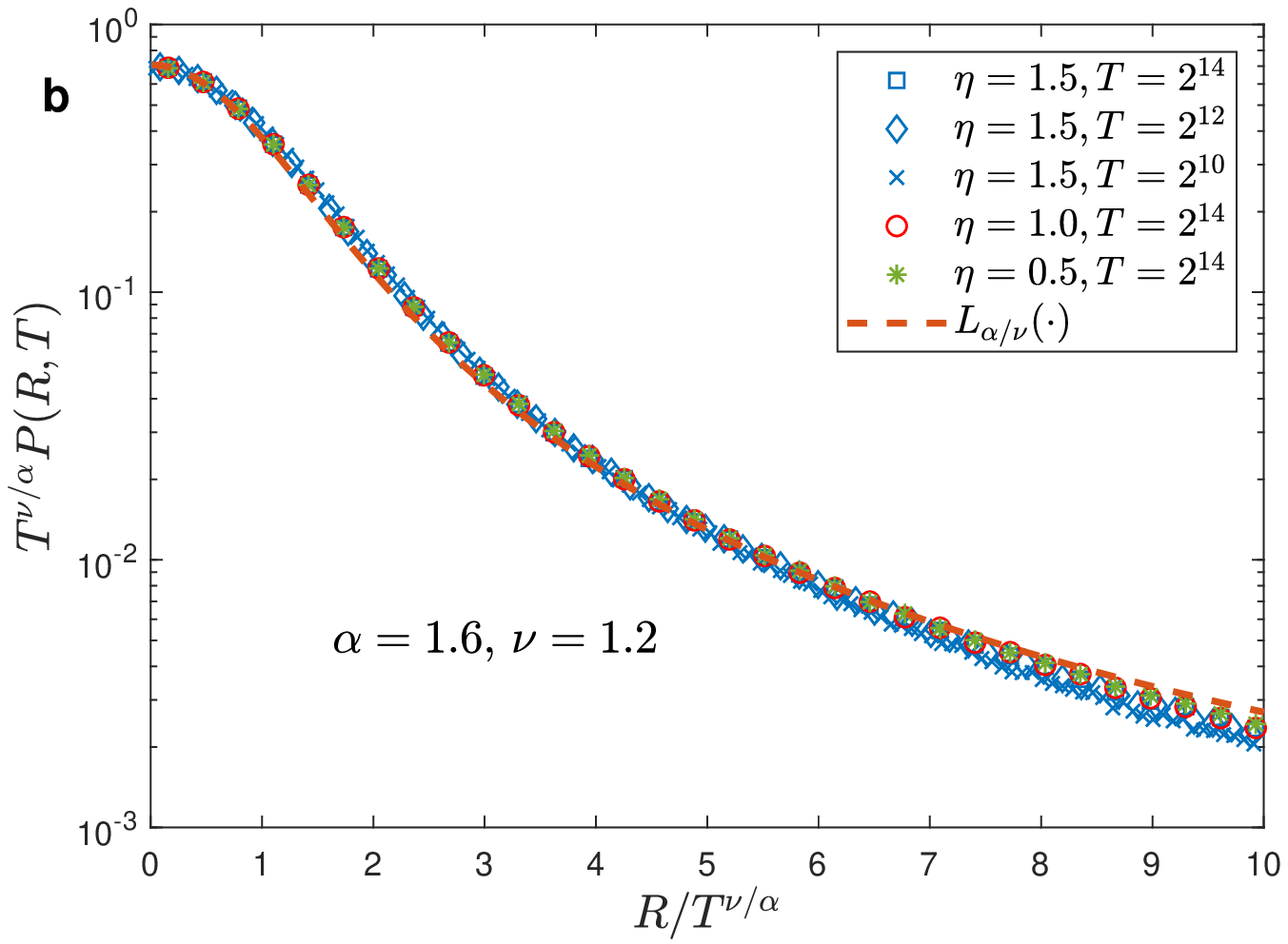}
\includegraphics[width=0.48\textwidth]{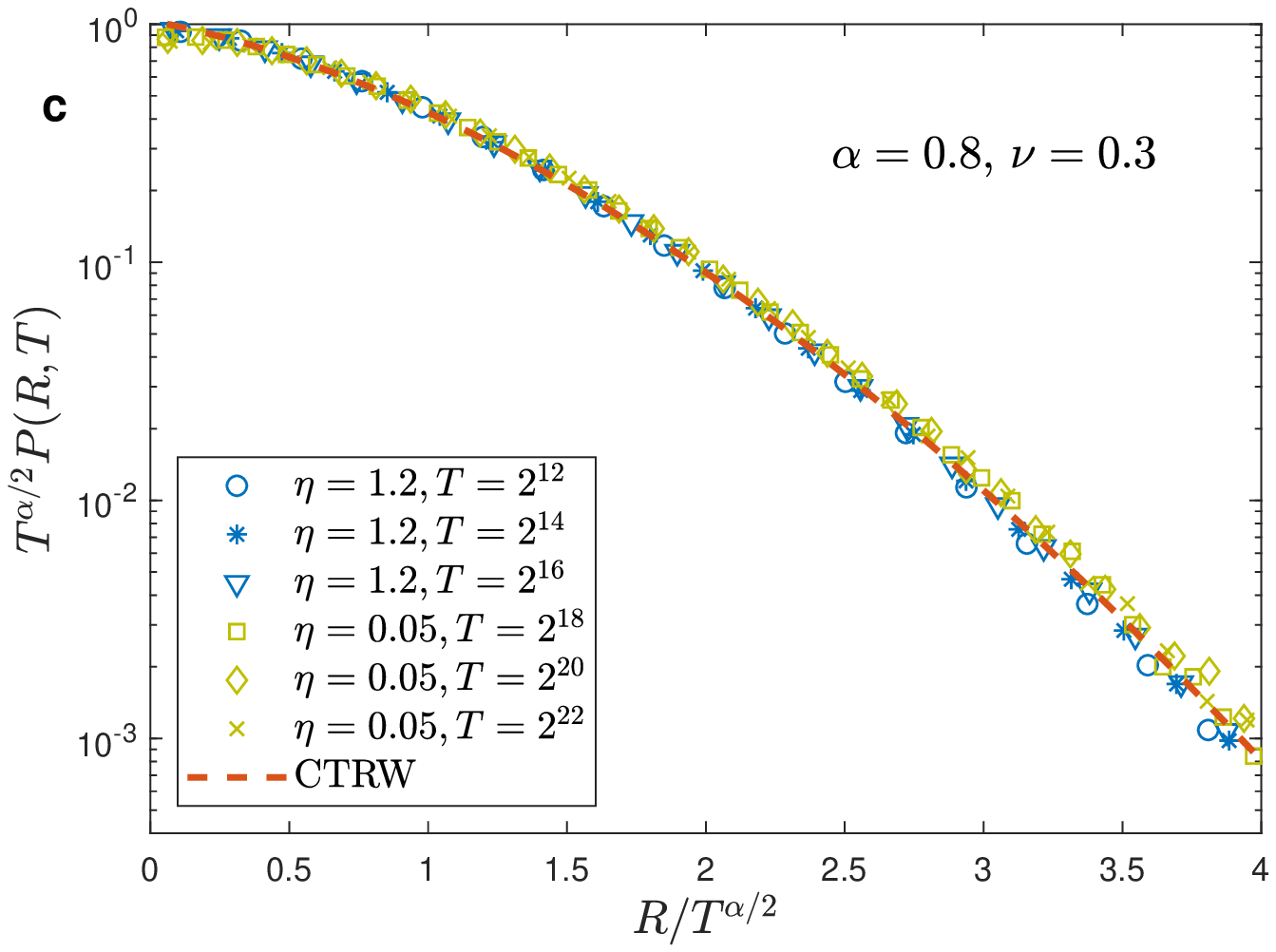}
\includegraphics[width=0.48\textwidth]{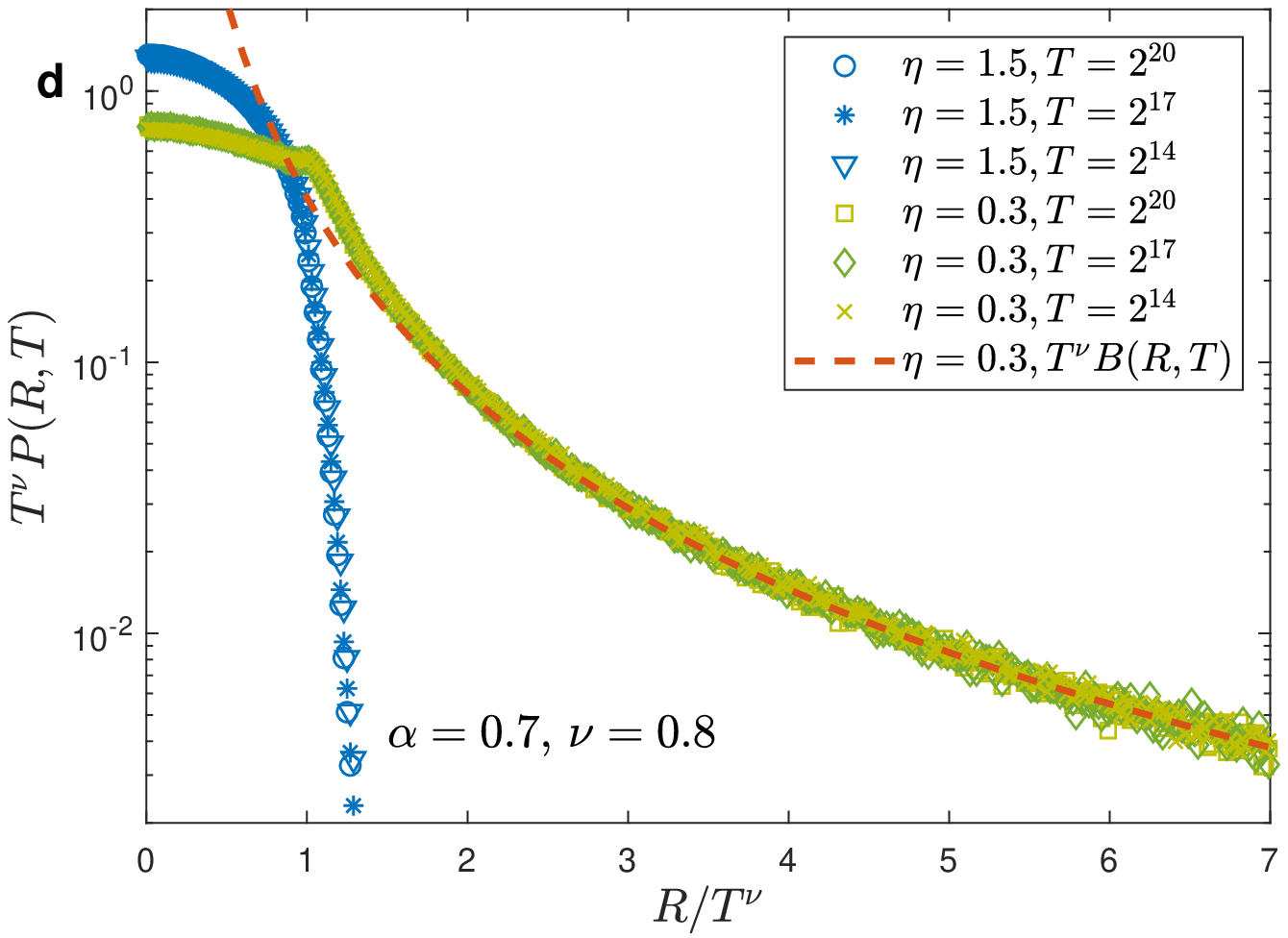}
	\caption{Scaling at short distances for the PDF in the generalized L\'evy walk model. In panel ({\bf a}) $\alpha=1.6>1$ and $\nu=0.7<\alpha/2$ we obtain a diffusive behavior with a Gaussian scaling function. In panel ({\bf b}) $\alpha=1.6>1$ and $\nu=0.7<\alpha/2$. Here the scaling length grows super-diffusively as $T^{\nu/\alpha}$  and the scaling function is the L\'evy function (see SI). In the panel ({\bf c}) $\alpha=0.8<1$ and $\nu=0.3<\alpha/2$, there is	sub-diffusion, the scaling length grows as $T^{\alpha/2}$ and the scaling function is the scaling function  of CTRW  with infinite waiting time, which is independent of $\nu$ (see SI). In panel ({\bf d}) $\alpha=0.7<1$ and $\nu=0.8>\alpha/2$, the scaling  described in SI  is determined by the single step motion $R\sim T^\nu$. The scaling function depends in a non-trivial way also on the exponents $\nu$ and $\eta$. The dashed line represents the result of the big jump approach in Formula (\ref{BJ3}). }
	\label{scalSHc}
\end{figure}
The bulk behavior of the PDF $P(R,T)$ can be evaluated showing that the following scaling form holds (see SI for details):
\begin{equation}
P(R,T)\sim \frac{f(R/\ell(T))}{\ell(T)}.
\label{scal}
\end{equation}
 with
\begin{equation}
{\ell(T)}\sim
\left\{
\begin{array}{ll}
\displaystyle
T^{1/2} &\mbox{if } \alpha>2 \nu \ \mbox{and }\ \alpha>1 \\
\displaystyle
T^{\nu/\alpha} &\mbox{if } \alpha<2 \nu \ \mbox{and }\ \alpha>1 \\
\displaystyle
T^{\alpha/2} &\mbox{if } \alpha>2 \nu \ \mbox{and }\ \alpha<1 \\
\displaystyle
T^{\nu} &\mbox{if } \alpha<2 \nu \ \mbox{and }\ \alpha<1 
\end{array}
\right.
\label{ell}
\end{equation}
For $\alpha> 2\nu$ and $\alpha>1$, the mean duration and the mean square length of the single step are finite so that the scaling function is Gaussian, independently of the value of the exponents $\alpha$, $\nu$ and $\eta$, as shown in Figure \ref{scalSHc} panel ({\bf a}). For $\alpha< 2\nu$ and $\alpha>1$ the mean duration of a step is finite but the mean square length is infinite, we are in a super-diffusive regime  and $f(\cdot)$ is a L\'evy stable function \cite{Chistyakov,Bouchaud} which only depends on the ratio $\nu/\alpha$, as shown in Figure \ref{scalSHc} panel ({\bf b}). Notice that in this case the exponent $\alpha/\nu$ driving both the scaling length $\ell(T)$ and the distribution $f(\cdot)$ is exactly the exponent that describes the distribution of the jump $L$ whose variance is infinite. For  
$\alpha> 2\nu$ and $\alpha<1$ the mean square length is finite but the mean duration of a step is infinite, and in this case the motion is sub-diffusive and $f(\cdot)$ only depends on $\alpha$ and corresponds to the scaling function of CTRW with infinite waiting time \cite{Bouchaud} (see Figure \ref{scalSHc} panel ({\bf c})). Finally, Figure \ref{scalSHc} panel ({\bf d}) shows that for $\alpha< 2\nu$ and $\alpha<1$, when  the mean square length and the mean duration are both infinite, 
the scaling function is not universal and depends on the exponents $\alpha$, $\nu$ and $\eta$.
In particular, the tail of the scaling function for $R/T^\nu\gg1$ is a pure power law when $\eta<\nu$ and in this case it can be evaluated using the big jump approach (dashed-line).

\subsubsection{Generalized L\'evy walks and the Big Jump: tails and rare events}

Let us now derive the tail $B(R,T)$ by applying the {big jump} principle. According to Eq. (\ref{BigJump}), we have to find the rate of attempts for the big jump, and the form of all the processes that, in a single jump, bring the walker in $R\gg\ell(T)$ at time T. We ignore the motion before and after the big jump, as this is the only contribution to the displacement.
As shown in Figure \ref{LWfig}, two possible different contributions to  ${\cal P}(R|T,t,T_w)$ are present. In panel ({\bf a})  $t>(T-T_w)$, the walker is still moving in the big jump at $T$ and $R=c t^{\nu-\eta}(T-T_w)^\eta$. In panel ({\bf b}) $t<(T-T_w)$, the walker ends its motion at $t$ so that $R=c t^\nu$. Since the big jump principle applies if $R\gg\ell(T)$, in this second process we get $c T^\nu \gtrsim c t^\nu = R\gg\ell(T)$. By comparing $\nu$ with the characteristic exponent of $\ell(T)$ in Eq. (\ref{ell}), we obtain that the path in panel ({\bf b}) is relevant only for $\alpha>1$ and $\nu>1/2$. On the other hand, in  the process of panel ({\bf a}) for $\nu>\eta$ the walker can reach arbitrary large distances in any fixed time interval $T-T_w$ and the process is always relevant. Finally, for $\nu\leq \eta$, for both processes in Figure \ref{LWfig}, we have that $R \sim cT^\nu$, and they both provide a contribution to ${\cal P}(R|T,t,T_w)$ only for $\alpha>1$ and $\nu>1/2$. 
This means that, for $\nu\leq \eta$, $\alpha<1$ and for  
$\nu\leq \eta$, $\alpha>1$, $\nu<1/2$ the walker cannot reach a distance larger than $\ell(T)$ in a single step and 
 Eq. (\ref{BigJump}) cannot be used to evaluate $B(R,T)$.

Let us first consider the case $\alpha>1$ and $\nu>1/2$ when both processes in Figure \ref{LWfig}
are relevant. In the SI we show that these processes can be simply encoded into the function ${\cal P}(R|T,L,T_w)$. Moreover since $\alpha>1$, the jump rate is constant ($n_R(T_w) =\langle t\rangle^{-1}$) and $p_{{\rm tot}}(t,T_w)=\lambda(t) /\langle t\rangle$. Then we plug $p_{{\rm tot}}(t,T_w)$ and the explicit expression for ${\cal P}(R|T,L,T_w)$ into the formula (\ref{BigJump}) so we obtain the explicit scaling form of $B(R,T)$:
\begin{equation}
B(R,T)=\frac{1 }{T^{\alpha-1+\nu}  }F\left(\frac{ R }{c T^{\nu} }\right)
\label{BJ1} 
\end{equation}
The scaling length at large distance grows as $cT^\nu$. The non universal scaling function $F(x)$ can be explicitly evaluated (see SI), it depends on the exponents $\alpha$, $\nu$ and $\eta$ and it is non-analytic at $x=1$.  
The case $\eta=\mu$  with $2\nu>\alpha$ has recently been studied in \cite{Aghion} and the far tails of the distribution have been obtained  using a moment summation technique.  The tail of standard L\'evy walks $\eta=\nu=1$ has been  discusses within various approaches \cite{Eli1,Wanli}.

\begin{figure}
\centering
\includegraphics[width=0.48\textwidth]{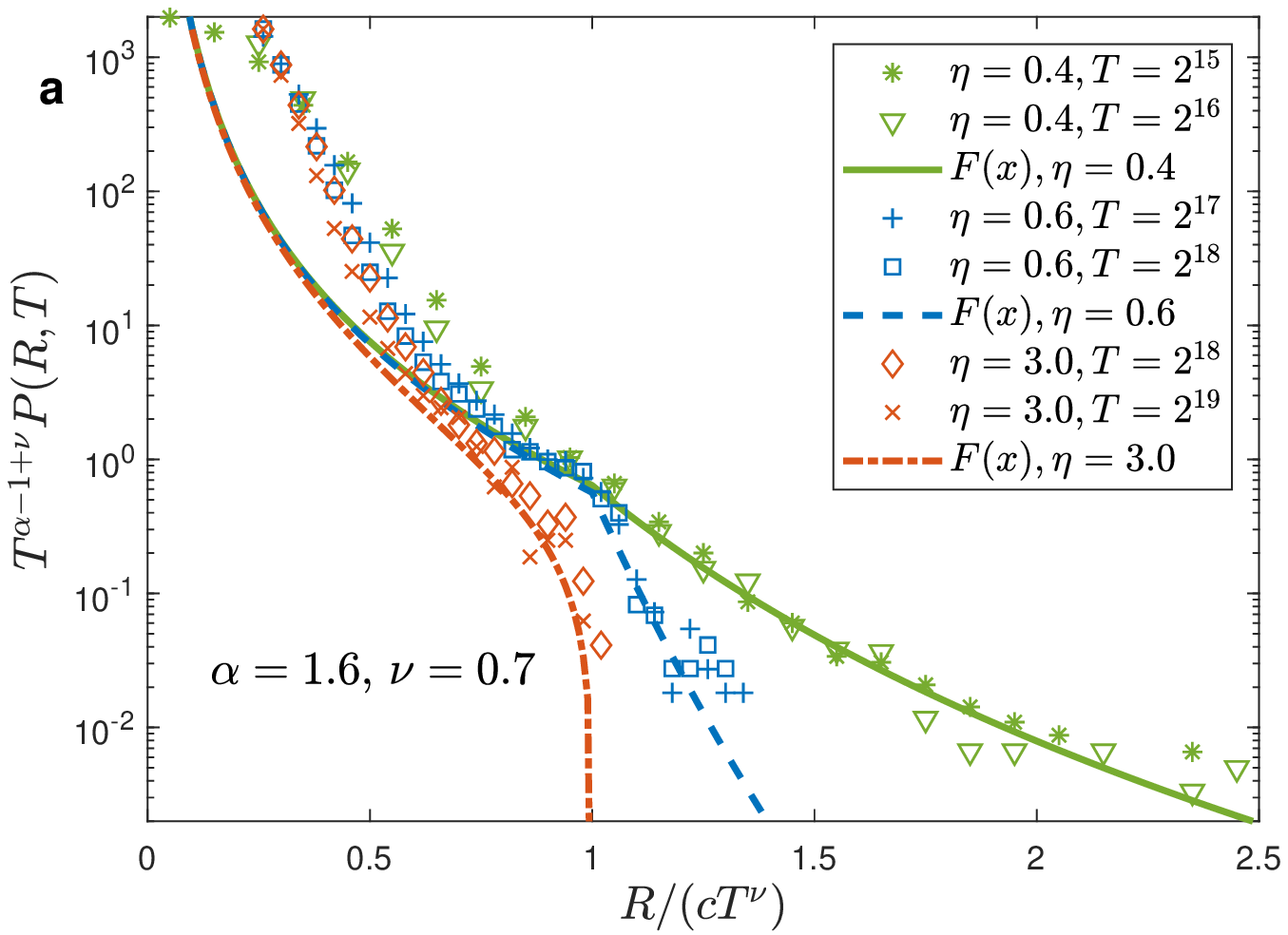}
	\includegraphics[width=0.48\textwidth]{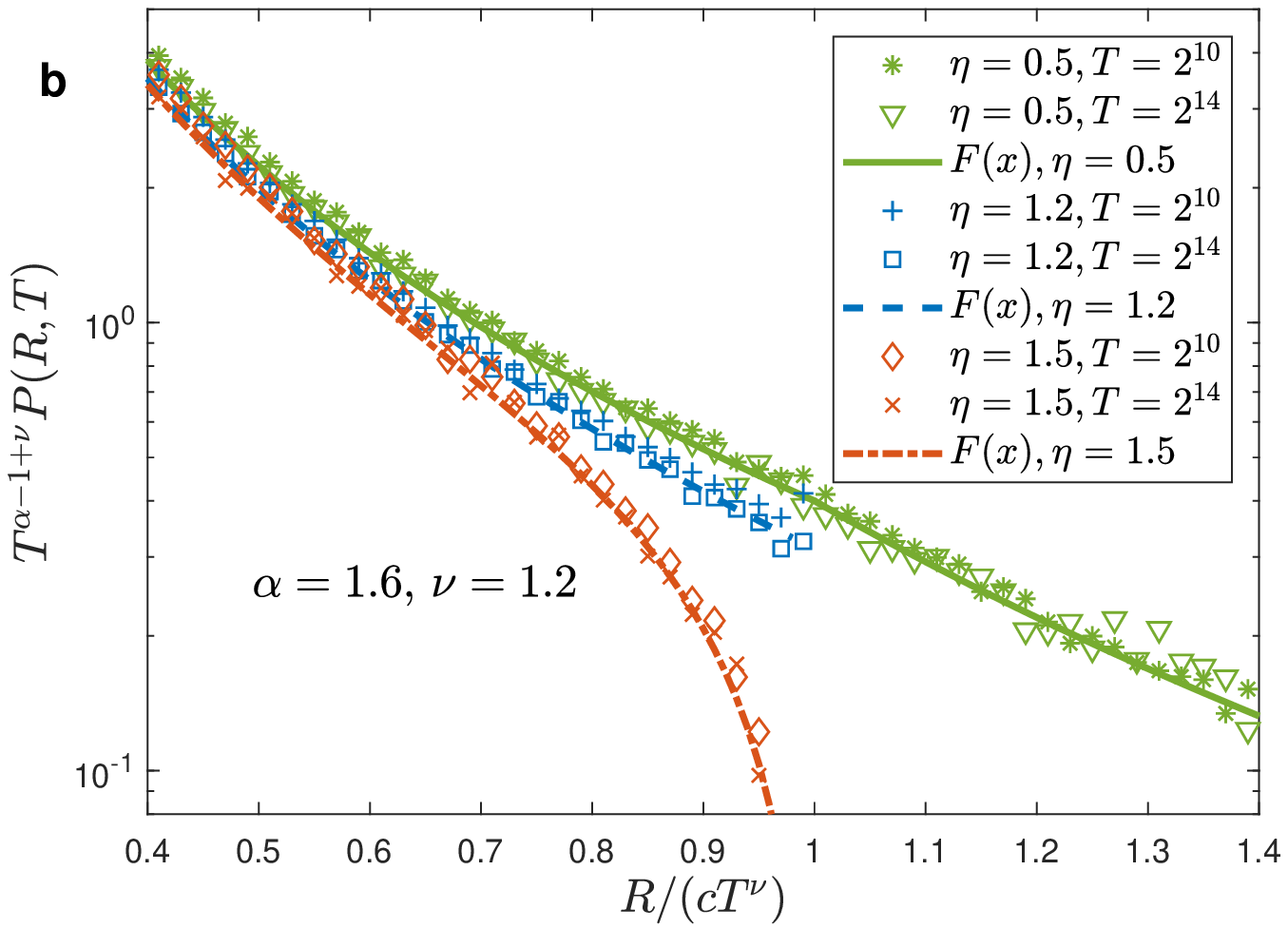}
	\caption{ The far tails of the distributions $P(R,T)$ for $\alpha=1.6>1$ and $\nu>1/2$: $\nu=0.7$ and $\nu=1.2$ in panel ({\bf a}) and ({\bf b}) respectively. The thick lines represent the theoretical value of the scaling function $F(x)$ explicitly calculated in the SI. The 
	plot shows the singular behavior of the scaling function when $x=R/T\nu=1$. Different behaviors are present  for $\eta<\nu$, $\eta>\nu$ and $\eta=\nu$ respectively.
For very small values of $\eta$ the cusp singularity in the distribution becomes barely visible.}
	\label{LW2fig}
\end{figure}

In Figure \ref{LW2fig}, panels ({\bf a}) and ({\bf b}), for $\alpha>1$ and $\nu>1/2$, we plot the far tail of $P(R,T)$ as a function of $R/(cT^\nu)$ and compare the analytic predictions with finite time simulations. In the long time limit, the densities fully agree with the big jump formalism. We remark that we used the same data of panels ({\bf a}) and ({\bf b})
in Figure (\ref{scalSHc}) introducing only a different scaling procedure.
In particular, the figure shows the singularities in the distribution when $R/(cT^\nu)=1$
and the different behaviors when $\nu>\eta$, $\nu=\eta$ and $\nu<\eta$ respectively.

In the case $\alpha>1$, $\nu<1/2$ and $\eta<\nu$ only the first process in Figure \ref{LWfig}
allows to reach distances larger than $\ell(T)$. Moreover since $\alpha>1$ and $\langle t \rangle$ is finite we have $p_{{\rm tot}}(t,T_w)=\lambda(t) /\langle t\rangle$. So we obtain (see SI):
\begin{equation}
B (R , T) =
\frac{ T^{\frac{\alpha\eta}{\nu-\eta}+1} c^{\frac{\alpha}{\nu-\eta}} \tau_0^\alpha }{\langle t \rangle (\nu+(\alpha-1)\eta)R^{1+\frac{\alpha}{\nu-\eta}}} 
 \label{BJ2} 
\end{equation}

Also for $\alpha<1$ and $\eta<\nu$ only the process in panel ({\bf a}) provides a contribution. For $\alpha<1$, however, the rate is not constant and $n_R(T_w) = C_\alpha T_w^{\alpha-1}/\tau_0^\alpha $  (the numerical constant $C_\alpha$ depends on $\alpha$ only) so we get (see SI):
\begin{equation}
B (R , T) =
\frac{ T^{\frac{\nu\alpha}{\nu-\eta}} c^{\frac{\alpha}{\nu-\eta}} D_\alpha }{ (\nu-\eta)R^{1+\frac{\alpha}{\nu-\eta}}} \label{BJ3} 
\end{equation}
where $D_\alpha$ depends on $\alpha$ only.
Since for $R\gg \ell(T)\gg cT^\nu$, no characteristic length is present in the system, Eqs. (\ref{BJ2}) and (\ref{BJ3}) are pure power-laws (scale free) functions decaying as $R^{-(1+\frac{\alpha}{\nu-\eta})}$.

\begin{figure}
\centering
\includegraphics[width=0.48\textwidth]{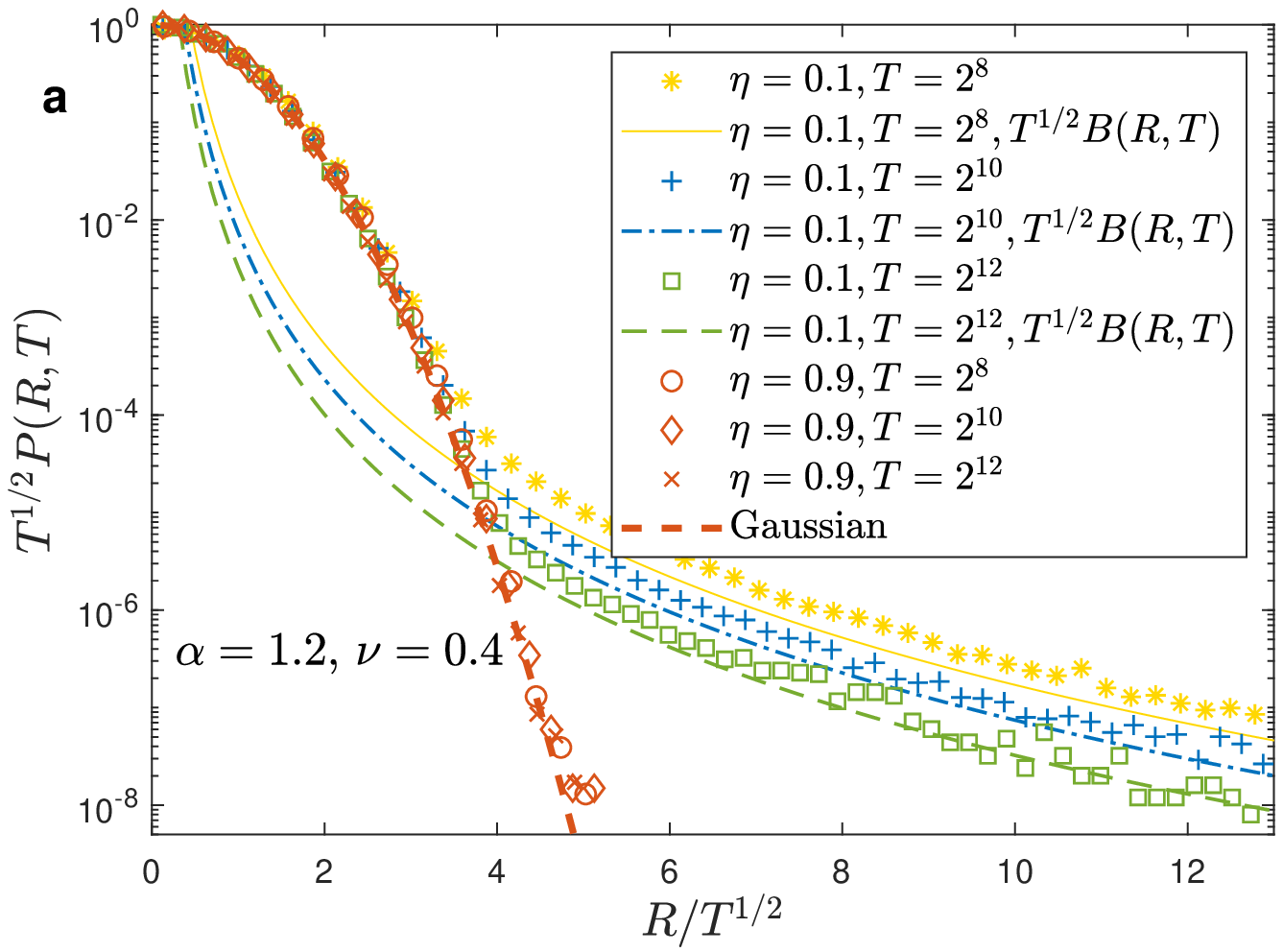}
	\includegraphics[width=0.48\textwidth]{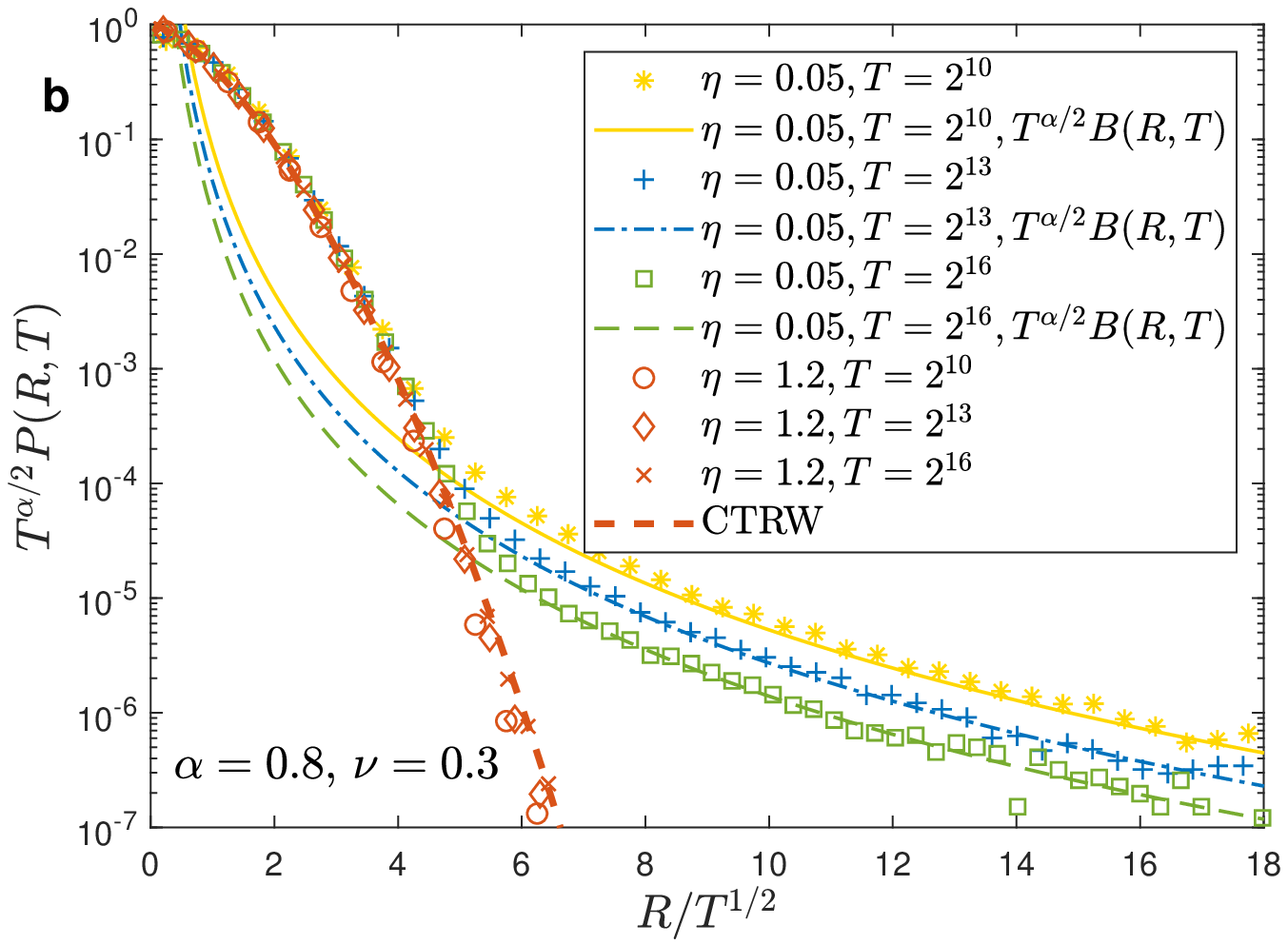}
	\caption{ Far tails of the distributions $P(R,T)$ for $\alpha=1.2>1$ and $\nu=0.4<0.5$ (panel ({\bf a})) and $\alpha=0.8<1$ $\nu=0.3<\alpha/2$ (panel ({\bf b})). The thick lines represent the big jump predictions when $\eta<\nu$ in formula (\ref{BJ2}) and (\ref{BJ3}) for the left and right panel respectively. The 
	plot shows the singular behavior of the scaling function when $x=R/T\nu=1$ and the different results when $\eta<\nu$, $\eta>\nu$ and $\eta=\nu$ respectively. For $\eta\geq\nu$ the figure shows that the bulk scaling function seems to describe the distribution even for $R>\ell(T)$. In this case, indeed, 
 Eq. (\ref{BigJump}) does not apply but the light cone grows much faster than $\ell(T)$. Therefore deviations at large distances are not given by a single process but by the contribution of many steps in the same direction, which is an exponentially suppressed process very difficult to be observed.}
	\label{LW3fig}
\end{figure}

Figure \ref{LW3fig}, panels ({\bf a}) and ({\bf b}), shows that, for $\alpha>1$, $\nu<1/2$ and for $\alpha<1$, $\nu<\alpha/2$,   Eq.s (\ref{BJ2}) and (\ref{BJ3}) well describe the distributions at $R\gg \ell(T)$, if $\eta<\nu$ (dashed-lines). In the regime, $\alpha<1$, $\nu>\alpha/2$, $\eta<\nu$ Figure \ref{scalSHc}, panel ({\bf d}), shows that the tail in Eq. (\ref{BJ3}) perfectly matches the short distance scaling function. Notice that in this last case Eq. (\ref{BJ3}) can be rewritten as in Eq.s (\ref{scal}) and (\ref{ell}) i.e. introducing the scaling length $\ell(T)\sim T^\nu$ and obtaining the same $T$ dependent pre-factor i.e. $B (R , T)\sim T^{-\nu} (R/T^\nu)^{-1-\frac{\alpha}{\nu-\eta}} $. This perfect matching means that for $\alpha<1$, $\nu>\alpha/2$ and $\eta<\nu$, Eq. (\ref{scal}) holds also for $R\gg \ell(T)$, however its behavior for $R\gg \ell(T)$ can be evaluated with the single big jump approach. 

For $\alpha<1$, $\eta\geq\nu$ and $\eta\geq\nu$,  $\alpha>1$ with  $\nu<1/2$ a single process cannot reach a distance larger than $\ell(T)$ and Eq. (\ref{BigJump}) does not apply. In particular the power law tails in Eq.s (\ref{BJ2}) and (\ref{BJ3}) cannot be observed, as shown in Figure \ref{scalSHc}, panel ({\bf d}), and in Figure \ref{LW3fig}, panels ({\bf a}) and ({\bf b}).   A summary of the scaling for the bulk and the tails in the whole range of exponents is shown in Table 1.

\begin{table}[]
\begin{tabular}{|c|l|l|l}
\hline
\multicolumn{1}{|l|}{}                                                        & \multicolumn{1}{c|}{$\nu>\alpha/2$}                                                                                                                                                 & \multicolumn{1}{c|}{$\alpha/2>\nu>1/2$}                                                                                                                                             & \multicolumn{1}{c|}{$1/2>\nu$}                                                                                                                                      \\ \hline
\begin{tabular}[c]{@{}c@{}}$\alpha>1$\\ Bulk\end{tabular}                  & \begin{tabular}[c]{@{}l@{}}Superdiffusion: \\ $\ell(T)\sim T^{\nu/\alpha}$\\ L\'evy Scaling\end{tabular}                                                             & \begin{tabular}[c]{@{}l@{}}Normal diffusion: \\ $\ell(T)\sim T^{1/2}$\\ Gaussian Scaling\end{tabular}                                                                            & \multicolumn{1}{l|}{\begin{tabular}[c]{@{}l@{}}Normal diffusion: \\ $\ell(T)\sim T^{1/2}$\\ Gaussian Scaling\end{tabular}}                                          \\ \hline
\multirow{2}{*}{\begin{tabular}[c]{@{}c@{}}$\alpha>1$\\ Tail\end{tabular}} & \begin{tabular}[c]{@{}l@{}}$\eta<\nu$: \\ $B(R,T)=\frac{1 }{T^{\alpha-1+\nu}  }F\left(\frac{ R }{c T^{\nu} }\right)$\\ $F(x)=x^{-1-\frac{\alpha}{\nu-\eta}}$ for $x>1$\end{tabular} & \begin{tabular}[c]{@{}l@{}}$\eta<\nu$: \\ $B(R,T)=\frac{1 }{T^{\alpha-1+\nu}  }F\left(\frac{ R }{c T^{\nu} }\right)$\\ $F(x)=x^{-1-\frac{\alpha}{\nu-\eta}}$ for $x>1$\end{tabular} & \multicolumn{1}{l|}{\begin{tabular}[c]{@{}l@{}}$\eta<\nu$: \\ $B(R,T)\sim \frac{ T^{\frac{\alpha\eta}{\nu-\eta}+1}  }{R^{1+\frac{\alpha}{\nu-\eta}}}$\end{tabular}} \\ \cline{2-4} 
                                                                              & \begin{tabular}[c]{@{}l@{}}$\eta\geq\nu$: \\ $B(R,T)=\frac{1 }{T^{\alpha-1+\nu}  }F\left(\frac{ R }{c T^{\nu} }\right)$ \\ $F(x)=0$ for $x>1$\end{tabular}                          & \begin{tabular}[c]{@{}l@{}}$\eta\geq\nu$: \\ $B(R,T)=\frac{1 }{T^{\alpha-1+\nu}  }F\left(\frac{ R }{c T^{\nu} }\right)$\\ $F(x)=0$ for $x>1$\end{tabular}                           & \multicolumn{1}{l|}{\begin{tabular}[c]{@{}l@{}}$\eta\geq \nu$: \\ Eq. (\ref{BigJump}) does not apply\end{tabular}}                                 \\ \hline
\multicolumn{1}{|l|}{}                                                        & \multicolumn{1}{c|}{$\nu>\alpha/2$}                                                                                                                                                 & \multicolumn{1}{c|}{$\alpha/2>\nu$}                                                                                                                                                 &                                                                                                                                                                     \\ \cline{1-3}
\begin{tabular}[c]{@{}c@{}}$\alpha<1$\\ Bulk\end{tabular}                     & \begin{tabular}[c]{@{}l@{}}Ballistic motion \\ $\ell(T)\sim T^\nu$\\ Non-universal scaling\end{tabular}                                                                             & \begin{tabular}[c]{@{}l@{}}Subdiffusion \\ $\ell(T)\sim T^{\alpha/2}$\\ CTRW scaling\end{tabular}                                                                                   &                                                                                                                                                                     \\ \cline{1-3}
\multirow{2}{*}{\begin{tabular}[c]{@{}c@{}}$\alpha<1$\\ Tail\end{tabular}}    & \begin{tabular}[c]{@{}l@{}}$\eta<\nu$: \\ $B(R,T)\sim \frac{ T^{\frac{\alpha\eta}{\nu-\eta}+1}  }{R^{1+\frac{\alpha}{\nu-\eta}}}$\\ Matching with the Bulk\end{tabular}                   & \begin{tabular}[c]{@{}l@{}}$\eta<\nu$: \\ $B(R,T)\sim \frac{ T^{\frac{\alpha\eta}{\nu-\eta}+1}  }{R^{1+\frac{\alpha}{\nu-\eta}}}$\end{tabular}                                            &                                                                                                                                                                     \\ \cline{2-3}
                                                                              & \begin{tabular}[c]{@{}l@{}}$\eta\geq \nu$: \\ Eq. (\ref{BigJump}) does not apply\end{tabular}                                                                      & \begin{tabular}[c]{@{}l@{}}$\eta\geq \nu$: \\ Eq. (\ref{BigJump}) does not apply\end{tabular}                                                                      &                                                                                                                                                                     \\ \cline{1-3}
\end{tabular}
\caption{A summary of the scaling behavior of bulk and tails for the PDF $P(R,T)$ when $\alpha>1$ and when $\alpha < 1$.}
\end{table}

We can compare the results of the tail in Table 1 with the conditions for the light cone.
For $\eta<\nu$ there is no light cone, so $B(R,T)$ describes the behavior of the tail at arbitrary large distances. When $\eta \geq \nu$ and $\alpha>1$ and $\nu> 1$, the tail $B(R,T)$ exactly vanishes at the light cone $l_{\rm cone}(T)=c T^\nu$. For $\alpha>1$ and $1/2 <\nu< 1$, 
$B(R,T)$ vanishes at $R=cT^\nu$. However in this case the particle can reach larger distances 
($l_{\rm cone}(T) \sim T$) with multiple steps. Clearly these processes are exponentially suppressed, and this means that  in the simulations of Figure \ref{LW2fig} panel ({\bf a}), for $\eta=3$ we observe events reaching a distance larger than $cT^\nu$ at time $T$ , but these events become extremely rare when increasing $T$.
When the big jump does not apply, two cases are possible: for $\alpha<1$ and $\nu\geq 1$, the light cone of the walker is determined by a single step, $l_{\rm cone}(T) = c T^\nu$, and trivially $B(R,T)=0$  since it is impossible to go farther than $c T^\nu \sim \ell(T)$ (as in the case of the standard L\'evy walks for $\alpha<1$ where $l_{\rm cone}(T) = c T$).
In the other cases, the light cone is reached in a large number of coherent steps all in the same direction and $l_{\rm cone}(T) \sim T \gg \ell(T)$,  whereas  the single jump cannot go farther than  $\ell(T)$. In this case we expect $B(R,T)$ to be exponentially suppressed and not described by Eq. (\ref{BigJump}).

\subsubsection{The Moments of the distribution}

\begin{figure}
\centering
	\includegraphics[width=0.48\textwidth]{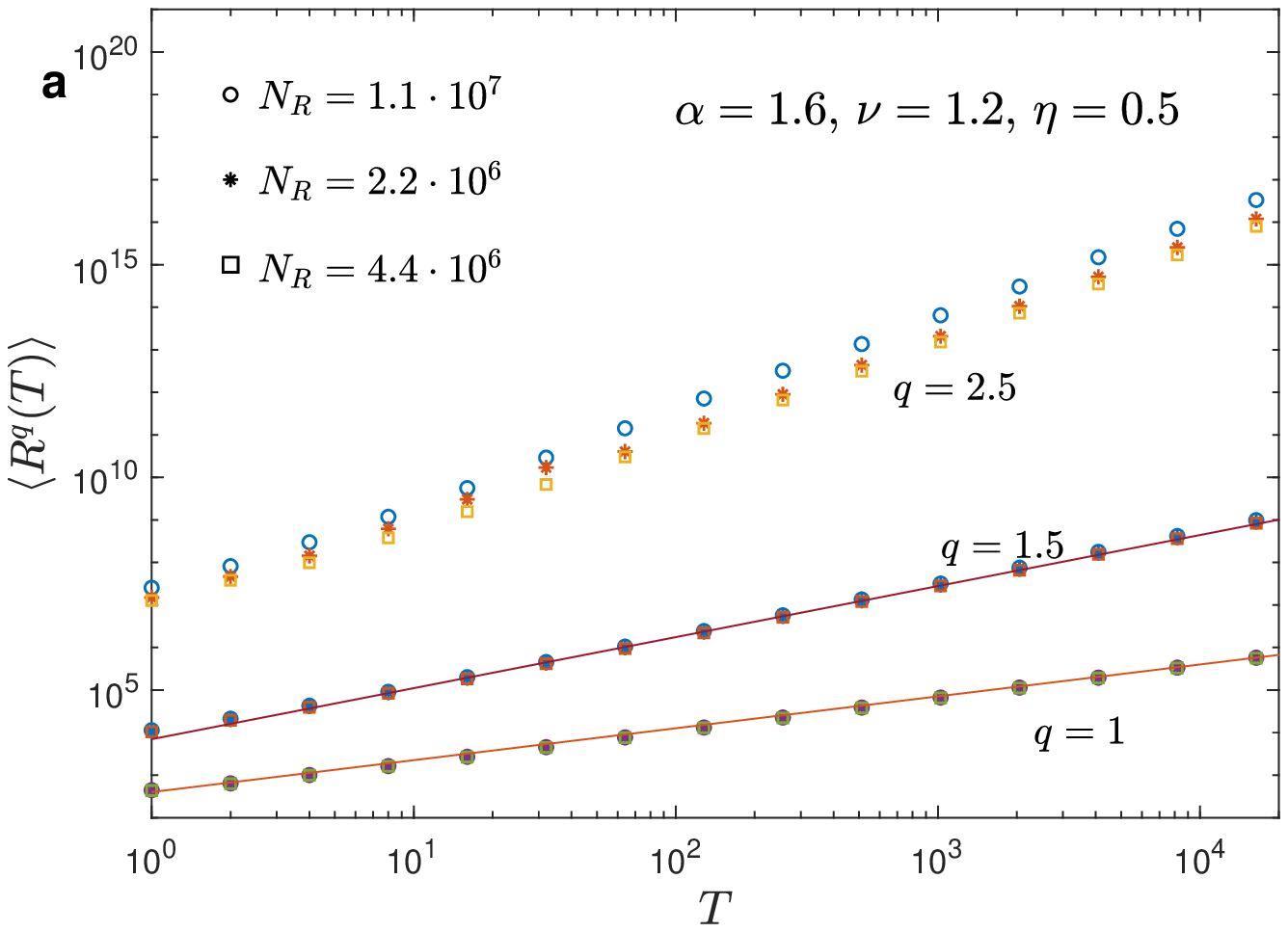}
	\includegraphics[width=0.48\textwidth]{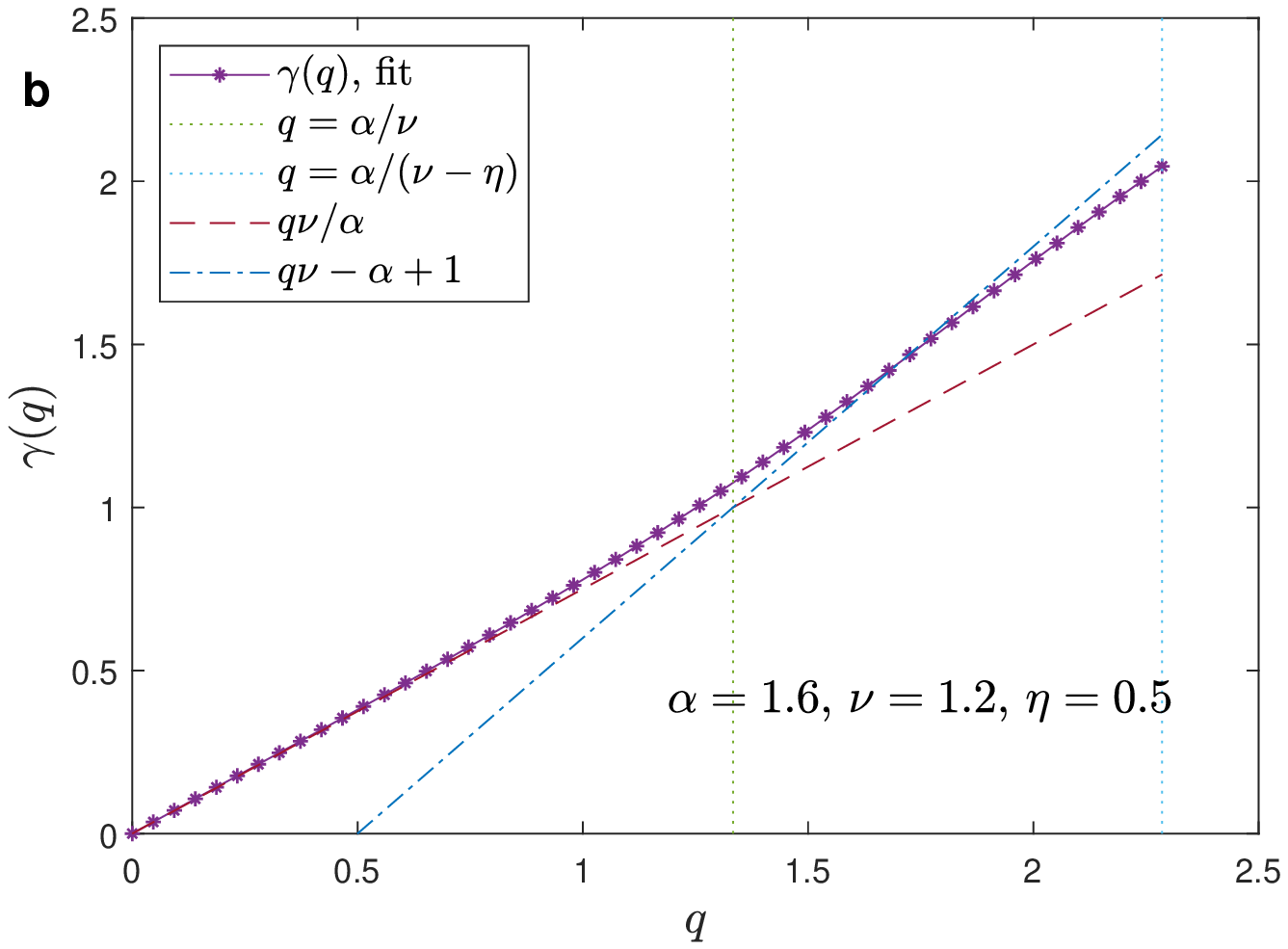}
	\caption{Moments of the distribution for $\alpha=1.6$, $\nu=1.2$ and $\eta=0.5$. Panel ({\bf a}): 
$\langle R^q(T)\rangle$ as a function of $T$ 
in the three regimes $q<\alpha/\nu$ ($q=1$), $\alpha/\nu<q< \alpha(\nu-\eta)$ ($q=1.5$) and $q> \alpha(\nu-\eta)$ ($q=2.5$).  Different symbols correspond to a different number of averages $N_R$. Continuous lines are the theoretical prediction $\langle R^q(T) \rangle \sim T^{\gamma(q)}$ according to Eq. (\ref{rP3}). In the first two regimes the symbols are perfectly superimposed and the results are independent of $N_R$. For $q> \alpha(\nu-\eta)$ instead the results depends on the number of realizations that we are averaging. In general, $\langle R^q(T)\rangle$ increases with $N_R$ but large fluctuations are present. In panel ({\bf b}) we plot the fitted exponent $\gamma$ as a function of the moment $q$. The three regimes $q<\alpha/\nu$, $\alpha/\nu<q< \alpha(\nu-\eta)$  and $q> \alpha(\nu-\eta)$ are shown. The theoretical result is well fitted but strong pre-asymptotic effects are present close to transitions points between the different regimes.  }
	\label{Momfig}
\end{figure}

\begin{figure}
\centering
	\includegraphics[width=0.48\textwidth]{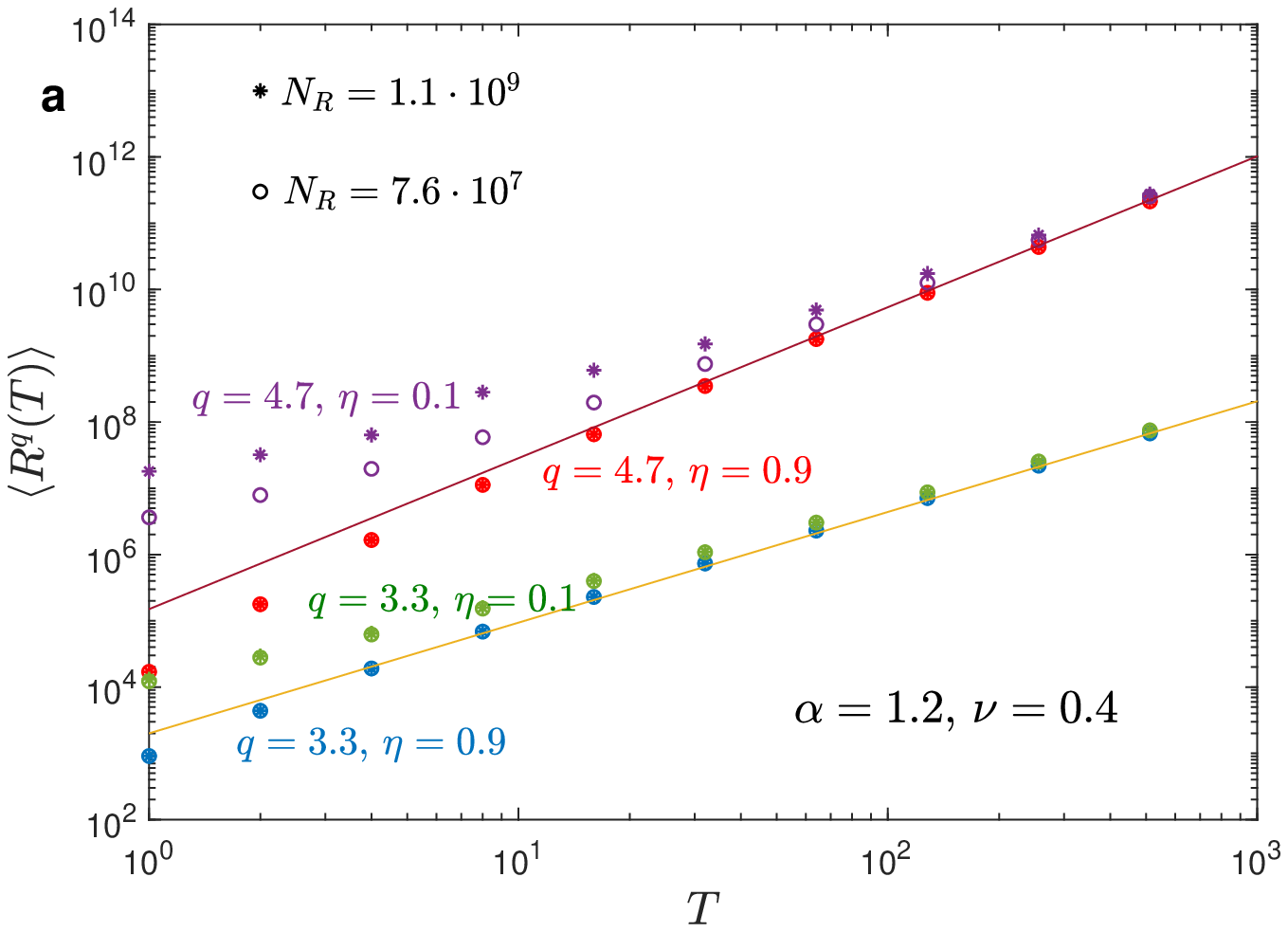}
	\includegraphics[width=0.48\textwidth]{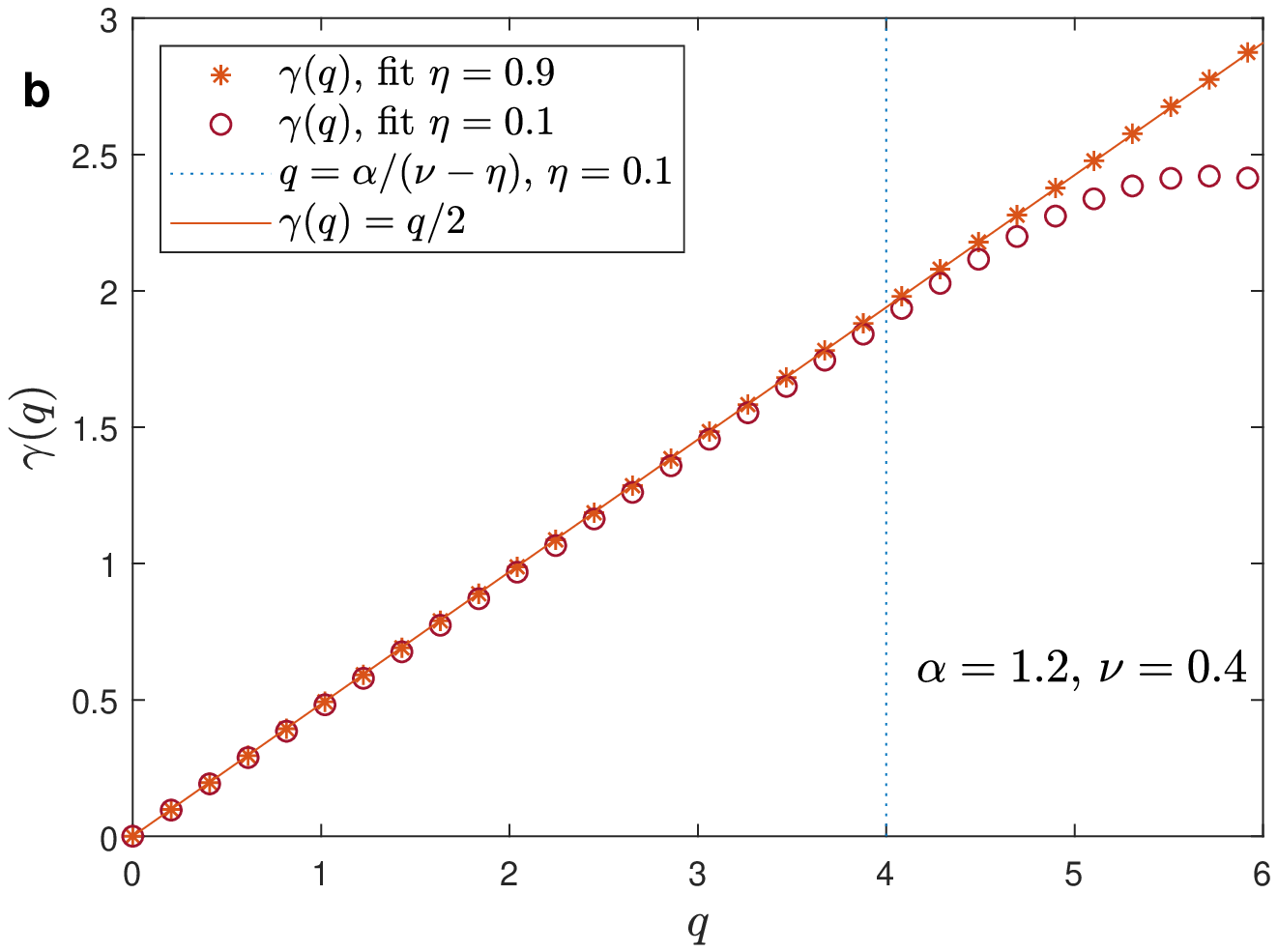}
	\caption{ Panel ({\bf a}): plot of
$\langle R^q(T)\rangle$ as a function of $T$ for $\alpha=1.2$, $\nu=0.4$, $\eta=0.1$ or $\eta=0.9$ and $q=4.7$ or $q=3.3$. For $\eta<\nu$ and $q> \alpha/(\nu-\eta)$ ($\eta=0.1$ and $q=4.7$) moments are infinite and simulations show a strong dependence on the number of dynamical realizations $N_R$. In the other cases the results are independent of $N_R$ and the theoretical results $\langle R^q(T)\rangle \sim T^{q/2}$ (continuous lines) asymptotically fit the simulations. Panel ({\bf b}): we plot the fitted exponent $\gamma$ as a function of the moment $q$.}
	\label{Momfig2}
\end{figure}

We now study the moments of the distribution of $R$, which are related to quantities typically measured in experiments. We introduce the exponents $\gamma(q)$ defined as 
$\langle R^q(T) \rangle \sim T^{\gamma(q)}$.  If $\gamma(q)$ is not simply proportional to $q$, this is what is called strongly anomalous diffusion \cite{castiglione,vollmer}. Here $\gamma(q)$ is evaluated taking into account the dominant term in Eq. (\ref{rP}) in the different regimes of Table 1.
Notice that, for $\eta<\nu$ $B(R,T)$ decays at large $T$ as $R^{-1-\frac{\alpha}{\nu-\eta}}$, therefore, for $q>\alpha/(\nu-\eta)$, the second integral in Eq. (\ref{rP}) and the relevant moments are infinite. In this case, as we show in Figs.
\ref{Momfig} and \ref{Momfig2}  the numerical value of $\langle R^q(T)\rangle$ depends on the number of realizations $N_R$ that we average in the simulation. In particular, $\langle R^q(T)\rangle$ 
diverges for $N_R\to \infty$ displaying at the same time very large fluctuations. 

In Figure \ref{Momfig} we consider the super-diffusive regime $\alpha>1$, $\nu>\alpha/2$ and $\nu>\eta$ where:
\begin{equation}
\gamma(q) = 
\left\{
\begin{array}{ll}
\displaystyle
  {q \nu / \alpha} &\mbox{if } q< \alpha/\nu \\
	\displaystyle
 {q \nu  -\alpha +1} &\mbox{if } \alpha/\nu<q<\alpha/(\nu-\eta)\\
\displaystyle
\infty &\mbox{if } q>\alpha/(\nu-\eta)
\end{array}
\right.
\label{rP3}
\end{equation}
Therefore the system displays strong anomalous diffusion \cite{castiglione,Cagnetta}.
In panel ({\bf a}) of Figure \ref{Momfig} we plot $\langle R^q(T)\rangle$ and we show
that when $\langle R^q(T)\rangle$ diverges, the results indeed depend 
on the number of realizations $N_R$ we use to obtain the average.
In panel ({\bf b}) we plot the function $\gamma(q)$ and we show that
far away from the critical value,  where preasymptotic effects are expected to be
stronger, simulations displays a nice agreements with theoretical values in Eq. \eqref{rP3}.

In Figure \ref{Momfig2}, panels ({\bf a}) and ({\bf b}), we consider $\alpha>1$ and $\nu<1/2$. For $\eta<\nu$ analytical calculations of  Eq. (\ref{rP}) gives  $\gamma(q)=q/2$ if $q<\alpha/(\nu-\eta)$ while $\gamma(q)$ diverges if $q>\alpha/(\nu-\eta)$. On the other hand, for $\eta\geq \nu$ we get $\gamma(q)=q/2$ for any values of $q$ and strong anomalous diffusion is not present. We remark that this is a general feature of the regimes where the big jump cannot be applied and the far tail are exponentially
suppressed. Figure \ref{Momfig2} confirms that simulations fit analytical predictions and that in the divergent regime the average moments depends on the number of dynamical realizations in the average process.

In general, therefore, the big jump approach via Eq. (\ref{rP}) is an effective tool for the calculations of anomalous exponents. Moreover, strong anomalous diffusion seems to be a general feature for systems where the big jump approach provides a significant contribution to the tail of $P(R,T)$. 

\section*{Discussion}
The single big jump principle provides an interesting and effective insight
on the origin of rare events in heavy - tailed processes. The principle allows both for
a physical interpretation of the mechanism that drives large fluctuations and 
also for a direct tool for calculation.  In practice, it works as soon as we deal with a
process where only one event contributes to the far tail, that is when only one jump takes our physical quantity $R$ to a value that is well beyond the scaling length of the process. 
While derived within a heuristic scheme, the principle in the rate approach appears to be extremely
effective in predicting the form of the tails, leaving an open question for a rigorous derivation.

We have here applied the principle to derive the exact form of the tail of the distribution in a class of generalized  L\'evy walks, a stochastic process that models anomalous transport in the presence of complex dynamics in the single step taken by the walker, which is subject to acceleration and deceleration effects. The dynamics in the steps give rise to a variety of shapes and behaviors for the PDF, summarized in Table 1. Interestingly, the single step dynamics is shown to strongly influence the form of the tail.
We are therefore in a situation where, while the bulk of the distribution feature the usual universality properties of central limit theorems, the tail is sensitive to the detail of the single step dynamics, because the single step is what drives the rare events.  

The big jump approach and the rate calculation can be applied well beyond the L\'evy walk models considered in this paper and well beyond quantities that represent random walkers, sums of steps and particle positions. Our result opens new possibilities to use rare events to obtain information on the microscopic dynamics and to have a fresh look on real datasets of single trajectories in systems exhibiting heavy tails statistics. In particular, we expect the generalized L\'evy walk to be largely applicable to all settings where deceleration and acceleration effects are relevant along the microscopic trajectories, like in contamination spreading  and in complex active transport in the cell \cite{future-levy,Gal}.

An open point  is to deal with processes where single rare events provide non trivial contribution to the distribution also at shorter distances \cite{Hoell}, as it happens in the case of the
standard L\'evy walk for $\alpha <1$. The extension of the results to higher dimensions \cite{fouxon} is also an open question.

\section*{Methods}

Consider a stochastic process where the variables $t_i$ ($i=1,2,\dots$) are drawn from the distribution $\lambda(t)$ at times $T_i$ with $T_i<T_j$ if $i<j$. The time $T_i$ is, in general,  a stochastic variable which can depend, according to the model, also on the draws occurring before $T_i$, i.e. on $t_1,\dots, t_{i-1}$.
A general expression for the PDF to measure the quantity $R$ at time $T$ is:
\begin{equation}
P(R,T)=\int \prod_i dt_i \lambda(t_i)  {\cal F}(R|T,\{t_i\})
\label{Pgen}
\end{equation}
where ${\cal F}(R|T,\{t_i\})$ is the probability of measuring $R$ at time $T$ given the sequence of random variables $\{t_i\}$.
    
Eq. (\ref{Pgen}) is very general and it
is suitable to describe processes with complex dynamical correlations, with  ${\cal F}(R|T,\{t_i\})$ being a highly non trivial function \cite{levyrand,VBB19,WVBB19}.

We first discuss the explicit form of  ${\cal F}(R|T,\{x_i\})$  for the generalized L\'evy walk
\cite{Albers,Sokolov}. We notice that only the first $n$ steps with $T_n<T<T_{n+1}$ provides a contribution to the process, so 
we can rewrite ${\cal F}(R|T,\{t_i\})$ as
\begin{equation}
{\cal F}(R|T,\{t_i\})=\sum_{n=1}^{\infty}  \theta(T-T_n) \theta(T_{n+1}-T) 
 \int \prod_{i=1}^n dc_i \frac{1}{2}(\delta(c_i-c)+\delta(c_i+c))\delta\left(R-\sum_{i=1}^{n-1} c_i t_i^\nu -c_n t_n^{\nu-\eta}(T-T_n)^\eta \right)
\end{equation}
where $\theta(\cdot)$ is the Heaviside function.
So we obtain for Eq. (\ref{Pgen}):
\begin{equation}
P(R,T)= \sum_{n=1}^{\infty} \int \prod_{i=1}^{i<n} d t_i \lambda(t_i) 
\theta(T-\sum_{i=1}^{n-1} t_i) \theta(\sum_{i=1}^{n} t_i-T) 
\int \prod_{i=1}^n d c_i \frac{1}{2}(\delta(c_i-c)+\delta(c_i+c))\delta\left(R-\sum_{i=1}^{n-1} c_i t_i^\nu -c_n t_n^{\nu-\eta}(T-\sum_{i=1}^{n-1} t_i)^\eta \right)
\label{Pgen_LW}
\end{equation}


Notice that in Eq. (\ref{Pgen_LW}) $P(R,T)$ is written as the sum of a series and each term of the series is given by an integral over a finite number $n$ of random variables. This is a general property since only processes occurring at time $T_n<T$ can affect the measure of quantity $R$ at time $T$. 
Let us consider again the general process in Eq. (\ref{Pgen}) where $t_i$ are generic random variables
drawn at times $T_i$. We can call $w_n(t_1,\dots,t_n,T)$ the probability that $T_n<T<T_{n+1}$ given the sequence of random variables $t_1,\dots,t_n$. Moreover we   
define ${\cal F}_n(R|T,t_1, \dots,t_n)$ the PDF to measure $R$ at time $T$ given the the random variables $t_1, \dots,t_n$ and knowing that the variables $t_n$ has been drawn before $T$ and the variable $t_{n+1}$ has been drawn after $T$.
We have
\begin{equation}
P(R,T)= \sum_{n=1}^{\infty} \int \prod_{i=1}^{i<n} d t_i \lambda(t_i)
w_n(t_1,\dots,t_n,T) {\cal F}_n(R|T,t_1, \dots,t_n)
\label{Pgen_2}
\end{equation}
comparing Eq. (\ref{Pgen_LW}) and Eq. (\ref{Pgen_2}) we have that $w_n(t_1,\dots,t_n,T)=
\theta(T-\sum_{i=1}^{n-1} t_i) \theta(\sum_{i=1}^{n} t_i-T) $ i.e. the probability is, respectively, zero or one  if the sums are smaller or larger than $T$ and 
\begin{equation}
{\cal F}_n(R|T,t_1, \dots,t_n) = \int \prod_{i=1}^n  d c_i \frac{1}{2}(\delta(c_i-c)+\delta(c_i+c))
\delta\left(R-\sum_{i=1}^{n-1} c_i t_i^\nu -c_n t_n^{\nu-\eta}(T-\sum_{i=1}^{n-1} t_i)^\eta \right)
\end{equation}
Moreover, we can now write a simple general definition of $\langle N(T) \rangle$ that is the average number of draws up to time $T$, i.e.
\begin{equation}
\langle N(T) \rangle = \sum_{n=1}^{\infty} n \int \prod_{i=1}^{i<n} d t_i \lambda(t_i)
w_n(t_1,\dots,t_n,T) 
\label{Pgen_3}
\end{equation}

Here, we have considered the generalized L\'evy walk and we provide a heuristic expression for ${\cal P}(R|T,t,T_w)$; analogous results have been obtained in \cite{VBB19} for different models such as the L\'evy Lorentz gas. A fundamental question for the stochastic process is to obtain a general procedure to obtain ${\cal P}(R|T,t,T_w)$ given the stochastic process described by the observable $R$ and the function ${\cal F}(R|T,\{t_i\})$ in Eq. (\ref{Pgen}).

\section*{Acknowledgements}

The support of Israel Science Foundation grant 1898/17
(E.B.) is acknowledged. R.B. thanks CSEIA (Center for Studies in European and International Affairs) of the University of Parma for the award granted to cover the costs of open-access publication.

\section*{Supplementary Information}
\subsubsection*{Short distance scaling}

Here we derive the bulk behavior of the PDF $P(R,T)$ for the generalized L\'evy Walk model, showing the non trivial dependence on the exponents $\alpha$ and $\nu$ described in Equations (6-7) of the main text.

Let us call $Q(R,T)$ the probability of making a jump at position $R$ and time $T$.
One can write:
\begin{equation}
Q(R,T) =  \delta(R)\delta(T)+\frac{1}{2}\int \left[Q(R-c t^\nu,T-t)  +Q(R+c t^\nu,T-t) \right]\lambda(t) dt 
\label{ap1}
\end{equation}
In the first term of the second member, the $\delta$-function takes into account that at time $T=0$ the walker is in $R=0$ and it makes a new step.

The probability $P(R,T)$ can be reconstructed from $Q(R,T)$ taking into account that a walker can arrive in $R$ only with a step of duration $t_2>t_1$ where $T-t_1$ is the time when $t_2$ have been extracted form $\lambda(t_2)$. We have:

\begin{equation}
P(R,T)  =   \int dt_1 \frac{1}{2}\left[Q(R-c t_2^{\nu-\eta} t_1^\eta,T-t_1)+Q(R+c t_2^{\nu-\eta} t_1^\eta,T-t_1) \right]\int_{t_1}^{\infty} dt_2 \lambda(t_2)
\label{ap2b}
\end{equation}

Let us consider $\tilde Q(k,s)$, i.e. the Fourier transform with respect $R$ and Laplace transform with respect $T$ of $Q(R,T)$. From Eq. (\ref{ap1}):
\begin{equation}
\tilde Q(k,s)   =  \frac{1}{1 - \tilde \lambda  (k,s)}
\label{ap3}
\end{equation}
where $\tilde \lambda(k,s)$ is:
\begin{equation}
\tilde \lambda(k,s)   =  \int dt \frac{\lambda(t)}{2}\left(e^{-st+ikct^\nu} + e^{-st-ikct^\nu} \right).
\label{ap3b}
\end{equation}
From Eq. \eqref{ap2b} we obtain the Laplace Fourier transform of $P(R,T)$
\begin{equation}
\tilde P(k,s)   =  \tilde Q(k,s) \tilde \gamma  (k,s)
\label{ap3c}
\end{equation}
where
\begin{equation}
\tilde \gamma(k,s)   =  \int_0^\infty dt_1 e^{-st_1} \int_{t_1}^\infty dt_2 \frac{\lambda(t')}{2}\left(e^{ikct_2^{\nu-\eta}t_1^\eta} + e^{-ikct_2^{\nu-\eta}t_1^\eta} \right).
\label{ap3d}
\end{equation}

Now we can expand $\tilde \lambda(k,s)$ and $\tilde \gamma(k,s)$ for small $s$ and $k$. Keeping only the leading terms in Eq. (\ref{ap3}) and (\ref{ap3}) for $\alpha>2\nu$ and $\alpha>1$ gives:
\begin{equation}
\tilde P(k,s) =  \frac{\langle t \rangle}{ s \langle t\rangle + (1/2) {k^2 \langle t^{2 \nu}\rangle}}.
\label{ap6}
\end{equation}
and $P(R,T)$ is a Gaussian centered in the origin 
\begin{equation}
P(R,T) =  \sqrt{\frac{\langle t \rangle}{2 \pi T c^2 \langle t^{2\nu} \rangle}} e^{-\frac {R^2 \langle t \rangle}{2 T c^2 \langle t^{2\nu} \rangle}}=\frac{1}{T^{1/2}} G\left(\frac{R}{T^{1/2}}\right).
\label{ap6b}
\end{equation}
where $G(\cdot)$ is a Gaussian scaling function and the characteristic length  of the process $\ell(T)$ grows as $\ell(T)\sim T^{1/2}$.
If $2 \nu> \alpha$ and $\alpha>1$ expanding $\tilde \lambda(k,s)$ and $\tilde \gamma(k,s)$ we obtain:
\begin{equation}
\tilde P(k,s) =  \frac{\langle t \rangle}{ s \langle t\rangle + A_\alpha {|k|^{\alpha/\nu} }}.
\label{ap7}
\end{equation}
where $A_\alpha=\int_0^\infty (1-\cos(t^\nu)) \frac{\tau_0^\alpha}{t^{1+\alpha}} dt$ is a constant. 
The inverse Fourier-Laplace transform gives:
\begin{equation}
P(R,T) = \frac{1}{T^{\nu/\alpha}} L_{\nu/\alpha}\left(\frac{R}{T^{\nu/\alpha}}\right).
\label{ap6b1}
\end{equation}
where  $L_{\nu/\alpha}(\cdot)$ is a L\'evy stable scaling function and the characteristic length now is  
$\ell(t)\sim t^{\nu/\alpha}$.
For $2 \nu < \alpha$ and $\alpha<1$ we obtain:
\begin{equation}
\tilde P(k,s) =  \frac{s^{\alpha-1} B_\alpha}{ s^\alpha B_\alpha + (1/2) {k^2 \langle t^{2\nu}\rangle}}.
\label{ap8}
\end{equation}
with $B_\alpha=\int_0^\infty (1-e^{-t}) \frac{\tau_0^\alpha}{t^{1+\alpha}} dt$. So the PDF scales as:
\begin{equation}
P(R,T)=\frac{1}{T^{\alpha/2}}C_{\alpha}\left(\frac{R}{T^{\alpha/2}}\right).
\label{ap8b}
\end{equation}
where $C_{\alpha}(\cdot)$ for $\alpha<1$ is the scaling function of CTRW with diverging average waiting times.
Finally for $ \alpha<2 \nu$ and $\alpha<1$ $\tilde \lambda(k,s)$ and $\tilde \gamma(k,s)$ cannot be expanded to the first order neither in $k$ or in $s$. In this case:
\begin{equation}
P(R,T)=\frac{1}{T^{\nu}}{f^*\left(\frac{R}{cT^{\nu}}\right)}.
\label{ap8b1}
\end{equation}
where $f^*(\cdot)$ is a non universal scaling function depending on $\alpha$, $\nu$, and $\mu$.

\subsubsection*{The Big Jump in Generalized L\'evy walks}
\label{MBJump}

Here, we perform the explicit calculation of the tail of the distribution $B(R,T)$ applying the big jump principle to the generalized L\'evy Walk model. The process has been described in Figure 1 of the main text.

Let us first consider the case $\alpha>1$ and $\nu>1/2$ where both processes in Figure 1 are relevant.
In particular, ${\cal P}(R|T,L,T_w)$ can be written 
introducing the Kronecker $\delta$-function and the Heaviside $\theta$-function:
\begin{equation}
{\cal P}(R|T,t,T_w)  =  \delta(R-ct^{\nu-\eta}(T-T_w)^\eta) \theta(t-(T-T_w)) + \delta(R-ct^\nu) \theta((T-T_w)-t)
\label{Lw1}
\end{equation}
where the first and the second terms correspond to the paths in panels ({\bf a}) and ({\bf b}) of Figure 1 respectively.
Since $\alpha>1$ the jump rate is constant ($n_R(T_w) =\langle t\rangle^{-1}$) and $p_{{\rm tot}}(t,T_w)=\lambda(t) /\langle t\rangle$ 
plugging Eq. (\ref{Lw1}) into formula (3) we obtain:
\begin{eqnarray}
B(R,T)
& = &B_0(R,T)+B_1(R,T)
\nonumber\\
B_0 (R , T) & = &\int_0^T \frac{dT_w}{\langle t \rangle} \int_0^\infty \frac{dt \tau_0^\alpha}{t^{1+\alpha}} \delta(R-ct^{\nu-\eta}(T-T_w)^\eta) \theta(t-(T-T_w)) , \label{Lw2}\\
B_1 (R,T) &= &\int_0^T \frac{dT_w}{\langle t \rangle} \int_0^\infty \frac{dt \tau_0^\alpha}{t^{1+\alpha}}
\delta(R-ct^\nu)\theta((T-T_w)-t)\nonumber
\end{eqnarray}
so that $B_0(R,T)$ and $B_1(R,T)$ are generated by the first and the second process illustrated in Figure 1 (main) respectively. Defining $y=ct^\nu$ in $B_1(R,T)$ we obtain:
\begin{eqnarray}
B_1(R,T)
& = & \int_0^T \frac{dT_w}{\langle t \rangle} \int_0^\infty \frac{dy  c^{\alpha/\nu} \tau_0^\alpha}{\nu y^{1+\alpha/\nu}}
\delta(R-y)\theta((T-T_w)-(y/c)^{1/\nu})=
\int_0^T \frac{dT_wc^{\alpha/\nu} \tau_0^\alpha }{\nu \langle t \rangle R^{1+\alpha/\nu}}  \theta((T-T_w)-(R/c)^{1/\nu})
\nonumber\\
 & = & 
\begin{cases} \frac{ c^{\alpha/\nu} \tau_0^\alpha  \left(T-\left(\frac{R}{c}\right)^{1/\nu}\right) }{\nu \langle t \rangle R^{1+\alpha/\nu}} =
\frac{ \tau_0^\alpha  \left(1-\left(\frac{R}{cT^\nu}\right)^{1/\nu}\right) }{ T^{\alpha-1+\nu} c \nu \langle t \rangle  \left(\frac{R}{cT^\nu}\right)^{1+\alpha/\nu}}
&\mbox{if } (R/c)^{1/\nu}<T \\
0 & \mbox{if } (R/c)^{1/\nu}>T  
\end{cases} \label{Lw3}
\end{eqnarray}

While in $B_0(R,T)$ we replace the integration variable $T_w$ with $T_2=T-T_w$ and then $t$ with
$y=ct^{\nu-\eta}T_2^\eta$, for $\eta<\nu$ we obtain:
\begin{eqnarray}
B_0 (R , T) & = &\int_0^T \frac{dT_2}{\langle t \rangle} \int_{T_2}^\infty \frac{dt \tau_0^\alpha}{t^{1+\alpha}} \delta(R-ct^{\nu-\eta} T_2^\eta)  
 = \int_0^T \frac{dT_2}{\langle t \rangle} \int_{c T_2^\nu}^\infty \frac{dy T_2^{\frac{\alpha\eta}{\nu-\eta}} c^{\frac{\alpha}{\nu-\eta}} \tau_0^\alpha}{(\nu-\eta)y^{1+\frac{\alpha}{\nu-\eta}}} \delta(R-y)  \nonumber\\
& = &\int_0^T \frac{dT_2}{\langle t \rangle}  \frac{ T_2^{\frac{\alpha\eta}{\nu-\eta}} c^{\frac{\alpha}{\nu-\eta}} \tau_0^\alpha}{(\nu-\eta)R^{1+\frac{\alpha}{\nu-\eta}}} \theta(R-c T_2^\nu) 
 = \int_0^{\min(T,(R/c)^{1/\nu})} dT_2 \frac{ T_2^{\alpha\eta/(\nu-\eta)} c^{\frac{\alpha}{\nu-\eta}} \tau_0^\alpha }{\langle t \rangle (\nu-\eta)R^{1+\frac{\alpha}{\nu-\eta}}}  \label{Lw4}\\
& = & 
\begin{cases} 
\frac{ c^{\alpha/\nu} \tau_0^\alpha \left(\frac{R}{c}\right)^{1/\nu} }{(\nu+(\alpha-1)\eta) \langle t \rangle R^{1+\alpha/\nu}} 
=
\frac{ \tau_0^\alpha \left(\frac{R}{c T^\nu}\right)^{1/\nu} }{T^{\alpha-1+\nu} c (\nu+(\alpha-1)\eta) \langle t \rangle \left(\frac{R}{c T^\nu}\right)^{1+\alpha/\nu}} 
&\mbox{if } (R/c)^{1/\nu}<T \\
\frac{ T^{\frac{\alpha\eta}{\nu-\eta}+1} c^{\frac{\alpha}{\nu-\eta}} \tau_0^\alpha }{\langle t \rangle (\nu+(\alpha-1)\eta)R^{1+\frac{\alpha}{\nu-\eta}}} 
=
\frac{ \tau_0^\alpha }{T^{\alpha-1+\nu} c   (\nu+(\alpha-1)\eta)\langle t \rangle \left(\frac{R}{c T^\nu}\right)^{1+\frac{\alpha}{\nu-\eta}}}
& \mbox{if } (R/c)^{1/\nu}>T  
\end{cases} \nonumber
\end{eqnarray}
While for $\eta>\nu$
\begin{eqnarray}
B_0 (R , T) & = &\int_0^T \frac{dT_2}{\langle t \rangle} \int_{T_2}^\infty \frac{dt \tau_0^\alpha}{t^{1+\alpha}} \delta(R-ct^{\nu-\eta} T_2^\eta)   = \int_0^T \frac{dT_2}{\langle t \rangle} \int_0^{c T_2^\nu} \frac{dy T_2^{\frac{\alpha\eta}{\nu-\eta}} c^{\frac{\alpha}{\nu-\eta}} \tau_0^\alpha}{(\eta-\nu)y^{1+\frac{\alpha}{\nu-\eta}}} \delta(R-y)  \nonumber\\
& = &\int_0^T \frac{dT_2}{\langle t \rangle}  \frac{ T_2^{\frac{\alpha\eta}{\nu-\eta}} c^{\frac{\alpha}{\nu-\eta}} \tau_0^\alpha}{(\eta-\nu)R^{1+\frac{\alpha}{\nu-\eta}}} \theta(c T_2^\nu-R) 
 = \int_{(R/c)^{1/\nu}}^{T} dT_2 \frac{ T_2^{\frac{\alpha\eta}{\nu-\eta}} c^{\frac{\alpha}{\nu-\eta}} \tau_0^\alpha }{\langle t \rangle (\eta-\nu)R^{1+\frac{\alpha}{\nu-\eta}}}  \label{Lw5}\\
& = & 
\begin{cases} 
\frac{ \tau_0^\alpha }{T^{\alpha-1+\nu} c (\nu+(\alpha-1)\eta) \langle t \rangle } \left( \frac{ \left(\frac{R}{ct^\nu}\right)^{1/\nu}  }{ \left(\frac{R}{ct^\nu}\right)^{1+\alpha/\nu}} - \frac{  1 }{\left(\frac{R}{ct^\nu}\right)^{1+\frac{\alpha}{\nu-\eta}}}\right)
&\mbox{if } (R/c)^{1/\nu}<T \\
0  & \mbox{if } (R/c)^{1/\nu}>T  
\end{cases} \nonumber
\end{eqnarray}
Finally for $\nu=\eta$:
\begin{eqnarray}
B_0 (R , T) & = &\int_0^T \frac{dT_2}{\langle t \rangle} \int_{T_2}^\infty \frac{dt \tau_0^\alpha}{t^{1+\alpha}} \delta(R-cT_2^\nu)  
 = \int_0^T \frac{dT_2}{\langle t \rangle} \frac{ \tau_0^\alpha}{\alpha T_2^{\alpha}} \delta(R-cT_2^\nu)
\nonumber\\
& = & \int_0^{cT^\nu} \frac{dy \left(\frac{y}{c}\right)^{1/\nu-1} \tau_0^\alpha }{c \alpha \langle t \rangle \alpha \nu \left(\frac{y}{c}\right)^{\alpha/\nu}} \delta(R-y) \label{Lw6}\\
& = & 
\begin{cases} 
\frac{  \tau_0^\alpha \left(\frac{R}{c}\right)^{1/\nu} }{\nu \alpha c \langle t \rangle \left(\frac{R}{c}\right)^{1+\alpha/\nu}} 
=
\frac{ \tau_0^\alpha \left(\frac{R}{c T^\nu}\right)^{1/\nu} }{T^{\alpha-1+\nu} c \nu \alpha \langle t \rangle \left(\frac{R}{c T^\nu}\right)^{1+\alpha/\nu}} 
&\mbox{if } (R/c)^{1/\nu}<T \\
0
& \mbox{if } (R/c)^{1/\nu}>T  
\end{cases} \nonumber
\end{eqnarray}
Eq.s (\ref{Lw2}-\ref{Lw6}) define the function $F(x)$ in Eq. (8). 
In particular, the non universal scaling function $F(x)$ depends on the exponents $\alpha$, $\nu$ and $\eta$ and it is non-analytic at $x=1$.  For $\eta\not= \nu$ $F(x)$ is continuous but non derivable at $x=1$. In particular, if $\eta>\nu$, $F(x)>0$ for all $x$. While, if $\eta<\nu$, $F(x)>0$ for $x<1$ and $F(x)=0$ for  $x>1$.  Finally, for $\eta=\nu$, $F(x)$ is discontinuous, dropping to $0$ at $x=1$ with $\lim_{x \to 1^-}F(x)>0$ and $F(x)=0$ for $x>1$.

For $\alpha>1$, $\nu<1/2$ and $\eta<\nu$ the jump rate is again constant and $p_{{\rm tot}}(t,T_w)=\lambda(t)/ \langle t\rangle$; however only the first process in Figure 1 (main) provide a contribution. In this case ${\cal P}(R|T,t,T_w)  =  \delta(R-ct^{\nu-\eta}(T-T_w)^\eta) \theta(t-(T-T_w))$ and we get the same result of Eq. \eqref{Lw4} (for $R>cT^\nu$), i.e. Eq. (9).

For $\alpha<1$ we have $p_{{\rm tot}}(t,T_w)=C_\alpha T_w^{\alpha-1}/t^{\alpha+1}$. Taking into account only of the first process in Figure 1 (main) for $\eta<\nu$ we have:
\begin{eqnarray}
B (R , T) & = & \int_0^T dT_w \int dt p_{{\rm tot}}(t,T_w)
\delta(R-ct^{\nu-\eta}(T-T_w)^\eta) \theta(t-(T-T_w))\nonumber\\
 & = &\int_0^T {dT_2} C_\alpha(T-T_2)^{\alpha-1} \int_{T_2}^\infty \frac{dt}{t^{1+\alpha}} \delta(R-ct^{\nu-\eta} T_2^\eta)  
  \nonumber\\
& = &\int_0^T \frac{dT_2 C_\alpha(T-T_2)^{\alpha-1}}{\nu-\eta} \int_{c T_2^\nu}^\infty \frac{dy T_2^{\frac{\alpha\eta}{\nu-\eta}} c^{\frac{\alpha}{\nu-\eta}} }{y^{1+\frac{\alpha}{\nu-\eta}}} \delta(R-y) 
= \int_0^T \frac{dT_2 C_\alpha(T-T_2)^{\alpha-1}T_2^{\frac{\alpha\eta}{\nu-\eta}} c^{\frac{\alpha}{\nu-\eta}} }{(\nu-\eta)R^{1+\frac{\alpha}{\nu-\eta}}}     \label{Lw8}\\
& = & 
\frac{ T^{\frac{\nu\alpha}{\nu-\eta}} c^{\frac{\alpha}{\nu-\eta}} C_\alpha \int_0^1 dx (1-x)^{\alpha-1} x^{\frac{\alpha\eta}{\nu-\eta}} }{ (\nu-\eta)R^{1+\frac{\alpha}{\nu-\eta}}} \nonumber
\end{eqnarray}
where $T_2=T-T_w$, $y=ct^{\nu-\eta} T_2^\eta$ $x=T_2/T$, and we use the fact that $R\gg \ell(T)\gtrsim c T^\nu>cT_2^\nu$ to fix $\delta(R-y)$. Setting $D_\alpha=C_\alpha \int_0^1 dx (1-x)^{\alpha-1} x^{\frac{\alpha\eta}{\nu-\eta}}$ we obtain the Eq. (10).

\end{document}